\documentclass{aa}  

\usepackage[usenames, dvipsnames]{xcolor}
\usepackage{graphicx}
\usepackage{txfonts}
\usepackage{amsmath}
\usepackage{textgreek}
\usepackage{natbib,twoopt}
\usepackage[colorlinks=true,allcolors=blue]{hyperref} 
\bibpunct{(}{)}{;}{a}{}{,} 
\makeatletter
\newcommandtwoopt{\citeads}[3][][]{\href{http://adsabs.harvard.edu/abs/#3}%
{\def\hyper@linkstart##1##2{}%
\let\hyper@linkend\@empty\citealp[#1][#2]{#3}}}
\newcommandtwoopt{\citepads}[3][][]{\href{http://adsabs.harvard.edu/abs/#3}%
{\def\hyper@linkstart##1##2{}%
\let\hyper@linkend\@empty\citep[#1][#2]{#3}}}
\newcommandtwoopt{\citetads}[3][][]{\href{http://adsabs.harvard.edu/abs/#3}%
{\def\hyper@linkstart##1##2{}%
\let\hyper@linkend\@empty\citet[#1][#2]{#3}}}
\newcommandtwoopt{\citeyearads}[3][][]%
{\href{http://adsabs.harvard.edu/abs/#3}
{\def\hyper@linkstart##1##2{}%
\let\hyper@linkend\@empty\citeyear[#1][#2]{#3}}}
\makeatother
\usepackage{siunitx}
\DeclareSIUnit \parsec {pc}
\DeclareSIUnit\angstrom{\text {Å}}
\DeclareSIUnit\year{\text {yr}}
\DeclareSIUnit\erg{\text {erg}}
\DeclareSIUnit\jansky{\text {Jy}}
\DeclareSIUnit \solarmass {\ensuremath{M_\odot}}
\DeclareSIUnit \h {\ensuremath{\mathit{h}}}

\usepackage[varvw]{newtxmath}
\renewcommand*\vec[1]{\ensuremath{\boldsymbol{#1}}}

\usepackage{orcidlink}

\usepackage{ulem}

\usepackage{comment}

\newcommand{\Msun}{M_\odot}

\begin{document} 

\title{Wandering and escaping:  recoiling massive black holes in cosmological simulations}
   \titlerunning{The cosmic effect of gravitational wave recoil}

   \author{Chi An Dong-P{\'a}ez\inst{1}\orcidlink{0000-0002-8590-4409}
          \and
          Marta Volonteri\inst{1}\orcidlink{0000-0002-3216-1322}
          \and 
          Yohan Dubois\inst{1}\orcidlink{0000-0003-0225-6387}
          \and 
          Ricarda S. Beckmann\inst{2}\orcidlink{0000-0002-2850-0192}
          \and
          Maxime Trebitsch\inst{3}\orcidlink{0000-0002-6849-5375}}

   \institute{Institut d’Astrophysique de Paris, UMR 7095, 
                CNRS and Sorbonne Universit\'{e}, 98 bis boulevard Arago, 75014 Paris, France\\
              \email{chiandongpaez@gmail.com}
        \and
            Institute for Astronomy, University of Edinburgh, Royal Observatory, Edinburgh EH9 3HJ, UK
        \and
            LERMA, Observatoire de Paris, PSL Research University, CNRS, Sorbonne Universit\'e, 75014 Paris, France
        }
        
   \date{Received ; accepted }

  \abstract{
  After a merger of two massive black holes (MBHs), the remnant receives a gravitational wave (GW) recoil kick that can have a strong effect on its future evolution. The magnitude of the kick ($v_\mathrm{recoil}$) depends on the mass ratio and the alignment of the spins and orbital angular momenta, therefore on the previous evolution of the MBHs. We investigate the cosmic effect of GW recoil by running for the first time a high-resolution cosmological simulation including GW recoil that depends on the MBH spins (evolved through accretion and mergers), masses and dynamics computed self-consistently. We also run a twin simulation without GW recoil. The simulations are run down to $z=4.4$.  We find that GW recoil  reduces the growth of merger remnants, and can have a significant effect on the MBH-galaxy correlations and the merger rate. We find large recoil kicks across all galaxy masses in the simulation, up to a few $10^{11}\,\si{\solarmass}$. The effect of recoil can be significant even if the MBHs are embedded in a rotationally supported gaseous structure. We investigate the dynamics of recoiling MBHs and find that MBHs remain in the centre of the host galaxy for low $v_\mathrm{recoil}/v_\mathrm{esc}$  and escape rapidly for high  $v_\mathrm{recoil}/v_\mathrm{esc}$. Only if $v_\mathrm{recoil}$ is comparable to $v_\mathrm{esc}$ the MBHs escape the central region of the galaxy but might remain as wandering MBHs until the end of the simulation. Recoiling MBHs are a significant fraction of the wandering MBH population. Although the dynamics of recoiling MBHs may be complex, some retain their initial radial orbits but are difficult to discern from other wandering MBHs on radial orbits. Others scatter with the halo substructure or circularise in the asymmetric potential. Our work highlights the importance of including GW recoil in cosmological simulation models.
  }

   \maketitle
%

\section{Introduction}

Massive black holes (MBHs) are found at the centre of most massive galaxies \citep{Kormendy1995,Magorrian1998}. MBHs co-evolve over cosmic history with their host galaxies as inferred observationally from the mass of the MBH ($M_\bullet$), found to correlate with the properties of its host galaxy, such as the mass and stellar velocity dispersion of the central bulge \citep{Magorrian1998,Tremaine2002,Marconi2003,Haring2004} and, with larger scatter, with the stellar mass of the galaxy ($M_\ast$) \citep{Reines2015,Greene2020}. Both the MBH and the stellar mass can grow efficiently from the dense galactic gas \citep[e.g.][]{DiMatteo2005,Springel2005,Dubois2014b}, which is replenished by cosmic inflows and cooling \citep[e.g.][]{White1978,Birnboim2003,Keres2005} and disrupted by the injection of energy from stars and MBH accretion themselves \citep[e.g.][]{Dubois2014a,Dubois2015,Habouzit2017,Angles-Alcazar2017}.

Mergers are another key ingredient in the co-evolution of MBH and their galactic hosts. The tidal forces from galaxy mergers can trigger a phase of rapid star formation and MBH accretion \citep{Hernquist1989,Barnes1991,McAlpine2018,Lapiner2021}. More importantly, a merger of two galaxies may lead to the merger of the two central MBHs, thus aggregating the masses of the two parent galaxies and MBHs in a correlated manner \citep{Peng2007,Kormendy2013}. The journey from a galaxy merger to an MBH merger is a complicated one. After two galaxies merge, the two central MBHs will dissipate energy to dynamical friction and sink to the centre of the new potential. If they form an MBH binary, the binary can continue to shrink until, if the evolution is fast enough, dissipation via gravitational waves (GWs) becomes efficient and the MBHs coalesce \citep{Begelman1980}. The future Laser Interferometer Space Antenna (LISA) mission aims to measure the GWs in the final stages of MBH mergers \citep{Amaro-Seoane2023}. 

In the last stage of coalescence, any asymmetries in the spins or masses of the binary lead to the final burst of GWs being emitted anisotropically and carrying away linear momentum. As a response, the remnant receives a velocity kick in the opposite direction \citep{Bekenstein1973}. The magnitude of this kick can reach values of up to $\SI{5000}{\kilo\meter\per\second}$ for MBHs with high spins partly aligned in the direction normal to the orbital plane but pointing in opposite directions in the orbital plane \citep{Lousto2011}. Even with other configurations, the velocity can reach several hundreds of ${\rm km\, s^{-1}}$. This means that the recoil velocity can exceed the escape velocity of haloes and displace MBH merger remnants from the nucleus or eject them from their hosts.

GW recoil can have a significant impact on the cosmic evolution of MBHs. The kicked MBH may see its growth severely reduced if it leaves the nuclear gas reservoirs and increases its velocity relative to the interstellar medium (ISM) until it dissipates its orbital energy and returns to the nucleus \citep{Blecha2008,Sijacki2011,Blecha2011,Blecha2016}. If accretion is reduced or the MBH is ejected from the nucleus, the energy injection into the nucleus is also reduced, which can assist nuclear star formation and deplete the gas available for future MBH growth \citep{Blecha2011}. For larger kicks that exceed the escape velocity of the host, the MBH can be ejected and the galaxy will be devoid of a central MBH until another MBH from another seeding event or galaxy merger is brought to the nucleus \citep{Schnittman2007,Volonteri2010,Dunn2020}. The MBHs that escape the centre can become `wandering' MBHs in the outskirts of the galaxy or in the galactic halo \citep{VPerna2005,Micic2006,Micic2011,Untzaga2024}

The reduction of MBH growth and the ejection of MBHs due to GW recoil can leave observational imprints. Ejected MBHs can reduce the MBH occupation fraction of galaxies, the MBH merger rate, and decrease the normalisation and increase the scatter of the MBH--host mass correlations \citep{Schnittman2007,Volonteri2007,Volonteri2010,Blecha2011,Volonteri2011,Dunn2020,Mannerkoski2022}. GW recoil also decreases the MBH merger rate \citep[e.g.][]{Micic2006,Dunn2020}. GW recoil can also leave a trace in the galactic morphology, acting in combination with MBH binary scouring to produce large stellar cores \citep{Gualandris2008,Nasim2021}. In some cases, the recoiling MBH could leave a trail of star-forming material \citep{Ogiya2024} or it could even be directly detected \citep{Komossa2012}.

Studying the cosmic effect of GW recoil is a difficult task. First, the magnitude of the kick, which regulates its effect on the future evolution of the MBH, is strongly dependent on the mass ratio, spins, and orbital angular momentum of the MBH pair. Extreme mass ratios or equal masses and aligned configurations of the angular momenta yield small kicks, while larger kicks require some degree of misalignment \citep[e.g.][]{Campanelli2007,Lousto2012}. These quantities depend complexly on the previous accretion, mergers and dynamics of the MBH pair \citep{DongPaez2023a}. Previous studies have found that high-density gas with coherent angular momentum aligns the orbital angular momentum and the spins leading to small recoil kicks, while, in dry mergers with little gas content, the directions are close to random and the recoil kicks can have a larger effect \citep{Bogdanovic2007,Dotti2010,Volonteri2010,Blecha2016,Dunn2020}. Moreover, the fate of a recoiling MBH depends on its dynamics in an often asymmetric and time-varying host potential in an expanding cosmology \citep{Sijacki2009,Sijacki2011,Blecha2011,Choksi2017}.

It follows that, to capture the cosmic effect of GW recoil and the complex co-dependence between GW recoil and MBH evolution and dynamics, one needs to run cosmological simulations that can capture the complexity of MBH environments and include realistic sub-grid models for accretion, spin evolution, and dynamics. Previous studies have used semi-analytic models relying on simplified prescriptions and often with unresolved dynamics, idealised simulations, or cosmological simulations that do not include all of these requirements. Often recoil kicks are estimated by assuming some distribution of spin magnitudes and directions instead of calculating them self-consistently.
So far the only cosmological hydrodynamical simulations modelling MBH evolution with GW recoil kicks on-the-fly are those of \cite{Sijacki2009} and \cite{Mannerkoski2022}. \cite{Sijacki2009} used a treatment of MBH spin evolution accounting for spin changes by mergers but neglected the role of MBH accretion for spin evolution, which is the dominant source of spin evolution for most grown MBHs~\citep{Dubois2014a,Bustamante2019,Peirani2024,Sala2024}.

In this paper, we present the first hydrodynamical cosmological simulation including GW recoil kicks calculated self-consistently from the masses, spins (evolved from both accretion and mergers), and dynamics. The masses and spins are evolved according to gas accretion and MBH mergers. The simulations also include sub-grid models for the unresolved dynamical friction onto the MBHs. We use this simulation to study the interplay between GW recoil and MBH evolution and the dynamics of recoiling MBHs in a cosmological environment. In section~\ref{sec:Method} we describe our simulations and our analysis method. In section~\ref{sec:MBH_evolution}, we study the evolution of MBH in our simulations. In section~\ref{sec:dynamics}, we study the dynamics of recoiling MBHs and compare recoiling MBHs to the global population of wandering MBHs. We conclude in section~\ref{sec:conclusions}.

\section{Method}
\label{sec:Method}

\subsection{The \textsc{Obelisk-Recoil} and \textsc{Obelisk-noRecoil} simulations}

We have run two cosmological zoom-in hydrodynamical simulations, \textsc{Obelisk-Recoil} and \textsc{Obelisk-noRecoil}, using the adaptive mesh refinement code \textsc{Ramses} \citep{Teyssier2002} down to $z\sim 4.4$. Our simulations are similar to the \textsc{Obelisk} simulation \citep{Trebitsch2021}, the main difference being that \textsc{Obelisk-Recoil} includes a sub-grid model for GW recoil kicks and both simulations add adjustments to other sub-grid models in \textsc{Obelisk}. We used these two simulations and \textsc{Obelisk} to cover a larger parameter space. Below we present a summary of the \textsc{Obelisk-Recoil} and \textsc{Obelisk-noRecoil} simulations, highlighting especially the differences with \textsc{Obelisk}. For a more detailed description of the simulation model, we refer the reader to \citet{Trebitsch2021}.

\subsubsection{Initial conditions}

The initial conditions are based on the \textsc{Horizon-AGN} simulation \citep{Dubois2014}. We selected particles enclosed within a sphere centred on the most massive halo in \textsc{Horizon-AGN} at $z=1.97$ with radius $2.51\,h^{-1}\,\mathrm{cMpc}$. We found the convex hull that encloses these particles in the initial conditions. This region was simulated with a succession of initial nested coarse grids down to maximum initial level of refinement $\ell=12$ corresponding to a high-resolution dark matter (DM) particle mass of $1.2\times10^6\,\rm M_\odot$. The rest of the $100\,h^{-1}\,\mathrm{cMpc}$ box was simulated at lower resolution with a minimum level of refinement $\ell=8$.

We assumed a standard $\Lambda$CDM cosmology. The cosmological parameters were taken to be the best-fit parameters from the WMAP-7 analysis \citep{Komatsu2011} -- Hubble constant $H_0 = 70.4\,\si{\kilo\meter\per\second\per\mega\parsec}$, dark energy density parameter $\Omega_\Lambda = 0.728$, total matter density parameter $\Omega_\mathrm{m} = 0.272$, baryon density parameter $\Omega_\mathrm{b} = 0.0455$, amplitude of the power spectrum $\sigma_8 = 0.81$, and spectral index $n_\mathrm{s} = 0.967$. 

\subsubsection{Gravity and hydrodynamics}

The gas was evolved using an unsplit second-order MUSCL--Hancock scheme \citep{vanLeer1979}. The inter-cell conservative variables were reconstructed using a total variation diminishing scheme with minmod slope limiter for the two states of the Riemann problem solved with the approximate Harten-Lax-van Leer-Contact Riemann solver. We assumed the gas is ideal and monoatomic with an adiabatic index of $5/3$. The time step follows the courant condition with a two-fold adaptive time stepping between levels larger than $\ell=13$ and single time stepping between coarser levels.

Any cell that exceeds a mass of 8 times the initial mass resolution is refined down to a minimum cell size of $\Delta x \simeq 35\,\si{\parsec}$ (proper, hence, corresponding to level $\ell=19$ and 20 when passing redshifts 9 and 4 respectively) within the zoomed-in region. DM, stars, and MBHs were modelled as collisionless particles, using a cloud-in-cell interpolation which size corresponded to that of the underlying cell (down to 35 proper pc), except for DM particles that has a minimum cloud size of 540 comoving parsec (i.e. level $\ell=18$ or 100 proper pc at $z=4.41$) in order to reduce the shot noise effect of the mass resolution imbalance between star and DM particles.
Gravitational acceleration is obtained solving the Poisson equation with a multigrid solver~\citep{Guillet2011} on levels coarser than $\ell<13$ and a conjugate gradient otherwise.
We further enforced the mesh to be refined down to the minimum cell size of $\Delta x\simeq 35\,\rm pc$ around MBHs to better capture their dynamics~\citep{Lupi2015}.

\subsubsection{Sub-grid modelling for gas and stars}

Unlike \textsc{Obelisk}, our simulations do not include radiation. Instead of using a non-equilibrium thermo-chemistry model for hydrogen and helium, we assume collisional-ionisation equilibrium in a homogeneous UV background~\citep{Haardt1996} below a redshift of reionisation of $z=8.5$, similarly to the \textsc{NewHorizon} simulation \citep{Dubois2021}. Metal-enriched gas can cool further following the cooling functions from \citet{Dalgarno1972} below a temperature of $T<10^4\,\rm K$, and \citet{Sutherland1993} for $T\ge 10^4\,\rm K$. We also account for the self-shielding against UV radiation of high-density gas \citep{Rosdahl2012}.

A cell was considered to be star-forming if the density of gas exceeds $5\,\mathrm{H}\,\si{\per\cubic\centi\meter}$ and the turbulent Mach number $\mathcal{M}\geq 2$. Star formation follows a \citet{Schmidt1959} law inversely proportional to the gas free-fall time with an efficiency parameter that depends on $\mathcal{M}$ and $\alpha_{\rm vir}=2E_{\rm k}/E_{\rm g}$ (where $E_{\rm k}$ and $E_{\rm g}$ are respectively the turbulent kinetic energy and the gravitational energy of the gas) following the multi-free fall model ~\citep{Hennebelle2011,Federrath2012} of \cite{Padoan2011}. Stars were modelled with $10^4\,\si{\solarmass}$ particles that represent stellar populations with a \citet{Kroupa2001} initial mass function. $5\,\si{\mega\year}$ after birth, $20\%$ of the stellar mass leads to supernova (SN) explosions, releasing metals with yield $y=0.075$ and an energy of $10^{51}\,\si{\erg}$ per $10\,\rm M_\odot$  individual SN into the gas, following the numerical scheme from \citet{Kimm2014}.

\subsubsection{Sub-grid modelling for MBHs}

We seeded MBHs with a mass of $M_\mathrm{seed}=10^5\,\rm M_\odot$ in cells where both the stellar and gas densities exceed a threshold of $n_\mathrm{gas}=200\,{\rm H}\,\si{\per\cubic\centi\meter}$. We note that our seeding model has a more stringent criterion and produces more massive seeds than in \textsc{Obelisk}, where $M_\mathrm{seed}=3\times10^4\,\rm M_\odot$ and $n_\mathrm{gas}=100\,{\rm H}\,\si{\per\cubic\centi\meter}$. To avoid the spurious formation of multiple seeds in one galaxy, we only allowed a cell to form a MBH if there are no other MBHs in a radius of $50\,\mathrm{ckpc}$.

The unresolved dynamical friction from collisionless particles and gas is modelled using the implementation by \citet{Dubois2013} and \citet{Pfister2019}. The frictional force from gas follows the analytical expression from \citet{Ostriker1999} but boosted by a factor $(\rho/\rho_\mathrm{DF,th})$ if $\rho>\rho_\mathrm{DF,th}=10\,\mathrm{H}\,\si{\per\centi\meter\cubed}$. We note that the normalisation of the dynamical friction force from the gas is over a factor of $10^6$ smaller than in \textsc{Obelisk}, arguably leading to more realistic MBH dynamics.

The MBHs in the simulation grow by gas accretion or MBH mergers. Gas accretion is modelled as spherical Bondi--Hoyle--Lyttleton accretion \citep{Hoyle1939,Bondi1944},
\begin{equation}
\label{eq:BHL}
    \dot{M}_\mathrm{acc}=4\pi G^2M_\bullet^2 \bar{B}.
\end{equation}
The variable $B$ is defined as $B=\rho/(c_\mathrm{s}^2+v_\mathrm{rel}^2)^{3/2}$, where $\rho$ is the gas density, $c_\mathrm{s}$ is the sound speed, and $v_\mathrm{rel}$ is the gas velocity relative to the MBH. The bar denotes the average in a Gaussian kernel around the MBH with a characteristic scale comparable to the BHL radius $R_\mathrm{BHL}=GM_\bullet/(c_s^2+v^2_\mathrm{rel})$.  We note that, compared to \textsc{Obelisk}, the average is computed on the composite quantity $B$ instead of the individual quantities $\rho$, $c_\mathrm{s}$, and $v_\mathrm{rel}$.

The accretion rate is limited by the Eddington rate,
\begin{equation}
\label{eq:Eddington}
    \dot{M}_\mathrm{Edd}=\frac{4\pi GM_\bullet m_\mathrm{p}}{\sigma_\mathrm{T}\varepsilon_\mathrm{r} c}\sim 2\,\si{\solarmass\per\year}\left(\frac{\varepsilon_\mathrm{r}}{0.1}\right)^{-1}\left(\frac{M_\bullet}{10^8\rm M_\odot}\right),
\end{equation}
where $m_\mathrm{p}$ is the proton mass, and $\sigma_\mathrm{T}$ is the Thompson cross-section, and $\varepsilon_\mathrm{r}$ is the radiated efficiency. The accretion flows are assumed to be radiatively efficient if $f_\mathrm{Edd}=\dot{M}_\mathrm{acc}/\dot{M}_\mathrm{Edd}\geq f_\mathrm{Edd,crit} = 0.01$, otherwise they are assumed to be fed by radiatively inefficient accretion flows (RIAFs). In the radiatively efficient regime, a fraction $\varepsilon_\mathrm{r}$ of the accreted mass is radiated and the remaining fraction is added to the MBH mass. The radiative efficiency $\varepsilon_\mathrm{r}$ depends on the MBH spin, ranging from $0.038$ for counter-rotating material to $0.32$ for co-rotating material around maximally spinning MBHs ($a_\bullet=\pm 0.998$). In the radiatively inefficient regime, the radiative efficiency is further decreased by a factor of $f_\mathrm{Edd}/f_\mathrm{Edd,crit}$ \citep{Benson2009}.

Every coarse time step of the simulation, two MBHs are allowed to merge if their separation becomes shorter than $4\Delta x$. Any losses to GWs are neglected, and the mass of the remnant is taken to be the sum of the two parent MBHs.

The spins of MBHs are evolved with gas accretion and MBH mergers following~\cite{Dubois2021} (similar to~\citealp{Dubois2014a, Dubois2014b} but with a modified model at $f_{\rm Edd}<f_\mathrm{Edd,crit}$). If accretion is radiatively efficient, the magnitude of the spin following an accretion episode is calculated as follows \citep{Bardeen1970},
\begin{equation}
\label{eq:spin_growth}
    a_\bullet=\frac{1}{3}\sqrt{\tilde r}\left(4-\sqrt{3\tilde r-2}\right),
\end{equation}
where $\tilde r=r_\mathrm{ISCO}[M_\bullet/(M_\bullet+\Delta M_\bullet)]^2$. The radius of the innermost stable circular orbit in units of the gravitational radius, $r_\mathrm{ISCO}$, is a function of the spin only. The spin magnitude evolved by accretion is limited to $-0.998<a_\bullet<0.998$ \citep{Thorne1974}. The angular momentum of the gas in the outer accretion disc is assumed to align with the gas on the resolved scales. The inner disc aligns or anti-aligns with the MBH spin through the \citet{Bardeen1975} effect depending on the criterion of \citet{King2005} for misalignement. The direction of the MBH spin following an accretion episode is calculated by summing the angular momenta of the accreted gas and the MBH. Assuming the MBH accretes from a reservoir of gas with coherent angular momentum, the MBH spin aligns with the gas in a short time scale of approximately \citep{Perego2009}
\begin{equation}
\label{eq:t_align}
    t_\mathrm{al}\sim 0.1 a_\bullet^{5/7} \left(\frac{M_\bullet}{10^6\rm M_\odot}\right)^{-2/35} f_\mathrm{Edd}^{-32/35}\,\mathrm{Myr}.
\end{equation}

In the RIAFs regime, the jets are powered from the rotational energy of the MBH, and the spin decreases following the simulations of \cite{McKinney2012} with the fitting function provided in~\cite{Dubois2021}.

Following a MBH merger the spin is updated as \citep{Rezzolla2008}
\begin{equation}
    \vec{a}_f = \frac{1}{(1+q)^2}\left(\vec{a}_1 +\vec{a}_2 q^2 +\vec{\ell} q \right),
    \label{eq:merger_spin}
\end{equation}
where $\vec{a}_1$ and $\vec{a}_2$ are the spins of the primary (the most massive) and secondary (least massive) MBHs. The mass ratio is the ratio of the secondary to the primary mass, $q=M_2/M_1$, and $\vec{\ell}$ is related to the orbital angular momentum, and calculated as
\begin{equation}
\begin{split}
\ell = & \frac{s_4}{\left(1+q^2\right)^2}\left(a_1^2 + a_2^2 q^4 + 2 \vec{a}_1 . \vec{a}_2 q^2\right)\\ 
& + \frac{s_5\eta + t_0 + 2}{1+q^2}\left(a_1 \cos \phi_1 + a_2 q^2 \cos \phi_2\right) \\
& + 2\sqrt{3} + t_2\eta + t_3\eta^2,
\end{split}
\end{equation}
where $\phi_1$ and $\phi_2$ are the angle of $\vec{a}_1$ and $\vec{a}_2$ compared to $\vec{\ell}$, and the symmetric mass ratio is $\eta = q/(1 + q)^2$. The constants are set to $s_4 = -0.129$, $s_5 = -0.384$, $t_0 = -2.686$, $t_2 = -3.454$, and $t_3 = 2.353$. $\vec{\ell}$ is taken to be aligned to the orbital angular momentum of the binary in the time step before the merger.

Active galactic nucleus (AGN) feedback was modelled using a dual-mode model. In the RIAF regime, the MBH injects kinetic energy in two cylindrical regions in the direction of its spin. This represents two relativistic jets of a radio-mode AGN. The fraction of energy into the jet is a function of the MBH spin and is calculated from a fit to the simulations of magnetically chocked accretion flows from \citet{McKinney2012}. In the radiatively efficient regime, the MBH injects as thermal energy $5\%$ of the accreted energy isotropically, similar to a radiative-mode AGN. We note that the feedback efficiency has been decreased compared to \textsc{Obelisk}, which assumes a fraction of $15\%$.

To assess our uncertainty in the accretion of MBHs, we run two additional simulations where the accretion rate is enhanced by setting $v_\mathrm{rel}=0$ in equation~\ref{eq:BHL}, one simulation with recoil and one without recoil. These simulations are run down to redshift $6$.

\subsubsection{Sub-grid GW recoil kicks following MBH mergers}

As described above, MBHs are merged if their separation decreases below $4\Delta x$. The remnant MBH is kicked in the coarse time step immediately after an MBH merger, which is an important simplification of the true MBH dynamics. Indeed, the final coalescence can take several hundred and up to several billions of years after they numerically connect at a scale of a few 100\,pc~\citep{Katz2020,Volonteri2020,Chen2022,Li2022,Li2024,DongPaez2023a}. However, we defer to future work the self-consistent treatment of MBH ``sub-grid'' dynamics (using e.g.~the RAMCOAL model of~\citealp{Li2024}) that leads to the final MBH binary coalescence and its GW recoil kick in a more consistent way.
We modelled the magnitude and direction of the recoil velocity using a fitting formula to the results of numerical relativity simulations, as quoted by \citet{Lousto2012}:
\begin{equation}
        \vec{v}_\mathrm{recoil}(q,\vec{a}_1, \vec{a}_2, \hat{\vec{L}}_\mathrm{orb})=v_m\vec{e}_1+v_\bot(\cos \xi \vec{e}_1 +\sin \xi \vec{e}_2) + v_\parallel \hat{\vec{L}}_\mathrm{orb}\, ,
\end{equation}
with
\begin{equation}
    \begin{aligned}
        &v_m=A_m\frac{\eta^2(1-q)}{1+q}\left[1+B_m\eta\right]\, ,\\
        &v_\perp=H\frac{\eta^2}{1+q}\left(a_{1,\parallel}-qa_{2,\parallel}\right)\, , \\
        &
        \begin{aligned}
            v_\parallel=&16\frac{\eta^2}{1+q}\left[V_{1,1}+V_A\tilde{S}_\parallel+V_B\tilde{S}_\parallel^2+V_C\tilde{S}_\parallel^3\right]\left|\vec{a}_{1,\perp}-q\vec{a}_{2,\perp}\right|\times\\
        &\times\cos\phi\, ,
        \end{aligned}
    \end{aligned}
\end{equation}
where $\vec{e}_1$ is the unit vector from the secondary, $\hat{\vec{L}}_\mathrm{orb}$ is the unit orbital angular momentum vector,  $\vec{e}_2=\hat{\vec{L}}_\mathrm{orb}\times\vec{e}_1$, $\eta=q/(1+q)^2$ is the symmetric mass ratio, and $\tilde{\vec{S}}$ is defined as
$\tilde{\vec{S}}=2(\vec{a}_1+q^2\vec{a}_2)/(1+q)^2$.
 The subscript $\parallel$ and $\perp$ refer to the parallel and perpendicular components relative to the orbital angular momentum. The constants are set to $\xi=145^{\circ}$, $H=6.9\times10^3\,\si{\kilo\meter\per\second}$, $A_m=1.2\times10^4\,\si{\kilo\meter\per\second}$, $B_m=-0.93$, $V_{1,1}=3677.76\,\si{\kilo\meter\per\second}$, $V_A=2481.21\,\si{\kilo\meter\per\second}$, $V_B=1792.45\,\si{\kilo\meter\per\second}$, $V_C=1506.52\,\si{\kilo\meter\per\second}$ \citep{Lousto2008,Lousto2009,Zlochower2011,Lousto2012}. $\phi$ is a phase we assume to be random following \citet{Lousto2011}. Recoil velocity is zero for equal-mass and equal-spin binaries, up to $\sim 200\,\si{\kilo\meter\per\second}$ for non-spinning black holes, and up to $\sim 5000\,\si{\kilo\meter\per\second}$ for spinning black holes in particular configurations.

The spins, masses, and orbital angular momentum of the primary and the secondary are measured at the time-step before the numerical merger and the resulting recoil velocity is injected into the merger remnant at the time-step after the MBH merger.

\subsection{MBH-galaxy assignment}

Galaxies and DM haloes were identified using the same method presented in \citet{Trebitsch2021}. We used a version of \textsc{AdaptaHOP} \citep{Aubert2004,Tweed2009} that runs on the total mass distribution of stars and DM. We only considered uncontaminated (with no low-resolution particles) galaxies and halos with more than $100$ star and DM particles. The centre of the galaxy was defined to be the peak of the stellar density distribution using a shrinking sphere approach \citep{Dubois2021}.

We assigned MBHs to galaxies using a similar approach to \citet{DongPaez2023a,DongPaez2023b}. We first assigned `main' MBHs, defined as the most massive MBH inside the half-mass radius $R_{50}$, to galaxies in descending order of stellar mass $M_\ast$. Unassigned MBHs are then assigned as `satellite' MBHs to the most massive galaxies that enclose them within $R_{90}$. $R_{90}$ is the radius encompassing $90\%$ of the stellar mass. Finally, any unassigned MBHs can be assigned as `halo' MBHs to the most massive galaxy that encloses them within the virial radius of their DM halo ($R_\mathrm{vir}$).

We assigned also MBH mergers to galaxies. Since recoil kicks may eject MBHs on time scales shorter than the typical time between two snapshots ($15\,\si{\mega\year}$), we assigned MBH mergers using the information from the snapshot before the merger. The host is taken to be the most massive galaxy hosting either of the MBHs within $r_{50}$. Mergers involving two MBHs located outside of $R_{50}$ are considered to be spurious and discarded from the sample. Nearly $30\%$ of mergers are discarded. 

\subsection{Analysis of recoiling and wandering MBH orbits}

The orbit of a recoiling MBH is followed after the kick. The position of the MBHs is recorded at every coarse time-step in the simulation (approximately every $\sim 0.1\,\si{\mega\year}$), while the position of each galaxy is obtained from the snapshots, recorded every $\sim 15\,\si{\mega\year}$. We obtain the MBH position relative to its host by interpolating the galaxy position between snapshots.

An MBH is considered to have returned to the centre of the galaxy after a recoil kick when there is an apocentre of its orbit that lies inside $R_{50}$. An MBH is considered to have escaped the halo if it escapes $R_\mathrm{vir}$. An MBH is considered to be escaping the halo if it is outside $R_{90}$ and its velocity is larger than four times the circular velocity at $R_\mathrm{vir}$. The factor of four is chosen by inspection of trajectories. An MBH is considered to be wandering if it remains inside the halo but has not returned to the centre either nor its velocity exceeds four times the circular velocity. See Table~\ref{tab:dynamical_classification} for a summary of this classification.

\begin{table*}[]
\caption{Summary of the dynamical classification for recoiling MBH that we use in this study.}
    \centering
    \small
    \begin{tabular}{l|l}
 \hline
        Returned MBH &  Apocentre is found inside $R_{50}$ \\
        Recoiling wandering MBH & Position inside $R_\mathrm{vir}$ but not in the other categories \\
        Escaping MBH & Position outside $R_{90}$ and inside $R_\mathrm{vir}$, and $v_\bullet>4v_\mathrm{circ,vir}$ \\
        Escaped MBH & Position outside of $R_\mathrm{vir}$\\
\hline
\end{tabular}
    \label{tab:dynamical_classification}
\end{table*}

The circularity of an orbit is defined as
\begin{equation}
\label{eq:circularity}
    \lambda=\frac{L_\bullet}{L_\mathrm{circ}(E_\bullet)}.
\end{equation}
$L_\bullet$ is the angular momentum of the MBH orbit with respect to the galaxy centre and $L_\mathrm{circ}(E_\bullet)$ is the angular momentum of the circular orbit with the same energy as the MBH orbit. The energy of the MBH orbit is calculated as $E_\bullet=0.5M_\bullet (\vec v_\bullet-\vec v_{\rm gal})^2 + \Phi(\vec x)$, where $\vec v_{\rm gal})$ is the velocity of the galaxy and $\Phi(\vec x)$ is the gravitational potential at the position of the MBH. We calculate the radius of the circular orbit with the same energy by spherically averaging the density around the centre of the galaxy.  $\lambda$ takes values from $0$ (radial orbits) to $1$ (circular orbits).

We have calculated the escape velocity of an MBH from its host halo as $v_\mathrm{esc}=\sqrt{-2[\Phi(\vec{x}_\bullet)-\Phi(\vec{R}_\mathrm{vir,rec})]}$. $\vec{R}_\mathrm{vir,rec}$ is the point in the surface of the virial radius in the direction of $\vec{v}_\mathrm{recoil}$. This is to make sure we account for any inhomogeneities on the $R_\mathrm{vir}$ surface. $\Phi(\vec{x})$ is obtained directly from the gravity solver of the simulation. For an MBH merger, $v_\mathrm{esc}$ is calculated from the snapshot before the merger. \citet{Blecha2011} show that the escape velocity can vary rapidly with time after a galaxy merger. We test for any time variation in $v_\mathrm{esc}$ by computing as an upper limit the value of $v_\mathrm{esc}$ from the centre of the galaxy after the merger. We find that the variation between our estimate and this upper limit is small -- most differences are below $10\%$ and the median difference is of $5\%$. This is probably because galaxy mergers in the simulations are more frequently minor mergers than major mergers, and therefore the potential is not dramatically deepened by the collision.

\section{GW recoil and the cosmic evolution of MBH}
\label{sec:MBH_evolution}

Throughout this study, we use the population of mergers from the \textsc{Obelisk-Recoil} and \textsc{Obelisk-noRecoil}. In addition to this, we also consider the \textsc{Obelisk} simulation. The \textsc{Obelisk} simulation includes a boosted sub-grid dynamical friction, that couples MBHs dynamically to the dense gas and increases the efficiency of accretion. We use \textsc{Obelisk} as a limiting case for efficient dynamical friction and accretion.

\begin{figure}
    \includegraphics[width=\columnwidth]{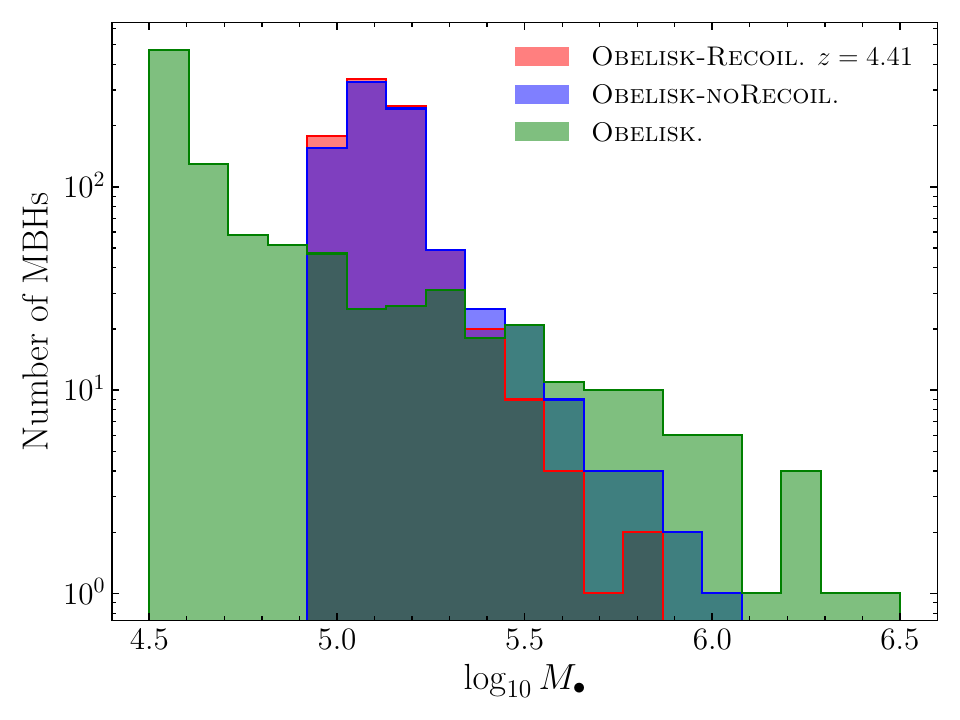}
    \caption{Mass of the main MBHs in the \textsc{Obelisk-Recoil} (red), \textsc{Obelisk-noRecoil} (blue) and \textsc{Obelisk} simulations at $z=4.41$.}
    \label{fig:main_BH_props}
\end{figure}

Figure~\ref{fig:main_BH_props} shows the MBH mass distribution in \textsc{Obelisk-Recoil} and \textsc{Obelisk-noRecoil}. The two simulations have similar mass distributions that only differ in the high-mass tail, where \textsc{Obelisk-Recoil} has a dearth of MBHs.

In the following section, we study the interplay between GW recoil and MBH evolution. GW recoil events are produced by MBH mergers. 

\subsection{Recoil and MBH growth}

\begin{figure}
    \includegraphics[width=\columnwidth]{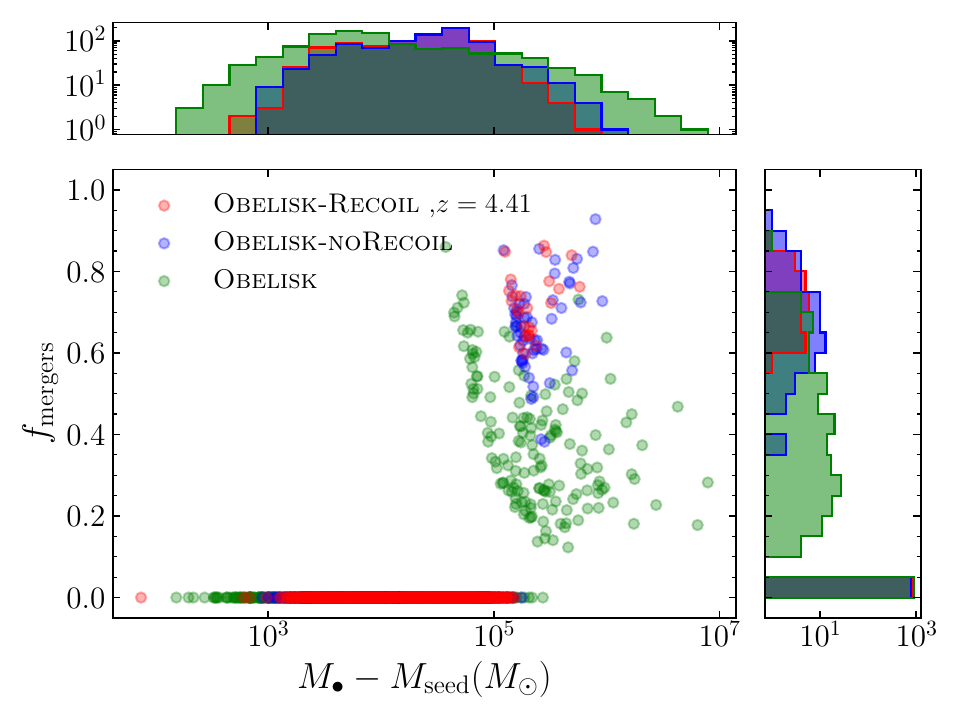}
    \caption{Fraction of mass gained through mergers against the total mass gained above the seed mass for \textsc{Obelisk-Recoil} (red), \textsc{Obelisk-noRecoil} (blue), and \textsc{Obelisk} (green).}
    \label{fig:BHgrowth_mergers}
\end{figure}

GW recoil can decrease the growth of MBH by ejecting and removing MBH mass from galaxies and by hindering gas accretion. In Fig.~\ref{fig:BHgrowth_mergers}, we show the fraction of the mass gained through mergers $f_\mathrm{mergers}$ as a fraction of the total mass gain, $M_\bullet-M_\mathrm{seed}$ for the three simulations analysed here, \textsc{Obelisk} (in green), \textsc{Obelisk-Recoil} (red), and \textsc{Obelisk-noRecoil} (blue), where $M_\mathrm{seed}$ is $3\times 10^4 \,\rm \Msun$ for the former and $10^5 \,\rm \Msun$ for the latter two. In \textsc{Obelisk-Recoil} and \textsc{Obelisk-noRecoil}, MBHs grow mostly through MBH-MBH mergers, while accretion is subdominant. In \textsc{Obelisk}, a significant fraction of the MBHs that have gained more than $10^5\,\rm \Msun$ settle in a phase of sustained efficient accretion. \textsc{Obelisk} features an \textit{ad hoc} boost in the gas dynamical friction, resulting in MBHs being much more dynamically coupled to the dense and cold ISM. As a result, MBHs are better centred in \textsc{Obelisk-Recoil} and its twin than in \textsc{Obelisk}, but they have a larger velocity relative to the gas, which suppresses the accretion rate.

\begin{figure}
    \includegraphics[width=\columnwidth]{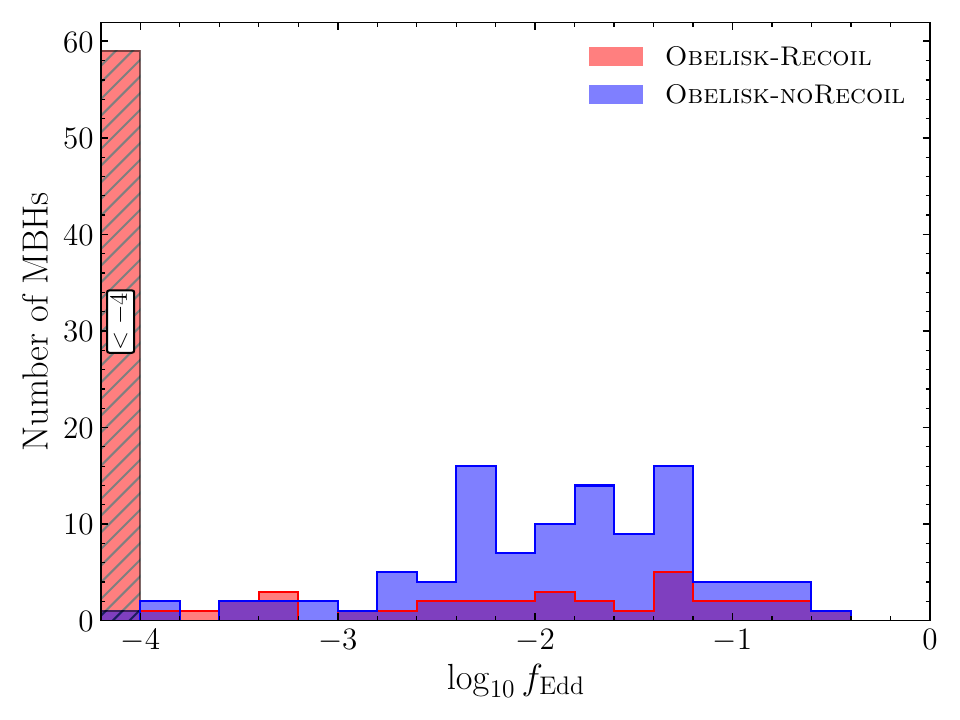}
    \caption{Distribution of the accretion rate normalised by the Eddington limit for \textsc{Obelisk-Recoil} and \textsc{Obelisk-noRecoil}, averaged over the $5$ Myr after the merger for the remnants. The first bin shows the number of MBHs below the lower limit of the plot.}
    \label{fig:fedd_mergers_rec}
\end{figure}

Recoil reduces the growth of MBHs both from accretion and mergers. The kick increases the relative velocity with the ISM, reducing the cross-section of the MBH for accretion. For stronger kicks, the MBH escapes the nucleus of the galaxy to a lower-density region, with less amount of gas to be potentially accreted. This effect is shown in Fig.~\ref{fig:fedd_mergers_rec}. The average accretion rate of remnant MBHs over the $5$ Myr following the mergers tends to experience a dramatic decrease after a recoil kick compared to the accretion rates of the parent MBHs before the merger. The accretion rates $5$ Myr following the merger in \textsc{Obelisk-Recoil} are significantly smaller than in \textsc{Obelisk-noRecoil}.
The MBH growth is suppressed until the MBH dissipates its kinetic energy and settles in the centre -- we discuss the timescales for an MBH to return to the centre of their host in section~\ref{subsubsec:escape}.

The integrated impact of the reduction in gas accretion over the cosmic evolution of MBHs is also seen in Fig.~\ref{fig:BHgrowth_mergers}. In \textsc{Obelisk-noRecoil} there are a few large MBHs ($M_\bullet-M_\mathrm{seed}>10^5\,\rm \Msun$) that have gained most of their mass through gas accretion, $f_\mathrm{mergers}<0.5$. In \textsc{Obelisk-Recoil} there are no MBHs in this regime.

The growth from mergers is also reduced. There are fewer MBH mergers in \textsc{Obelisk-Recoil} (92 mergers) compared to \textsc{Obelisk-noRecoil} (104 mergers): this is because recoil can remove MBHs from the centre of galaxies, decreasing the MBH occupation probability of galaxies.
In total, there are $13$ fewer mergers, a difference of $\sim10\%$. We note that the number of mergers for the simulation with recoil is possibly overestimated since in our simulations new MBHs can be seeded if the original MBH has been ejected from a galaxy and there are no other satellite MBHs. 

The total effect of recoil on the growth of the global MBH population is explored in Fig.~\ref{fig:main_BH_props}. The difference in MBH mass between the two simulations is noticeable. The \textsc{Obelisk-Recoil} lacks the high mass tail present in the \textsc{Obelisk-noRecoil}. The distributions in the host galaxy mass and $f_\mathrm{Edd}$ (not shown) are similar for the two simulations. Recoil does not have a significant effect on the growth of the global galaxy population. This is partly because the MBHs do not grow efficiently in our simulation and they do not release significant energy into the ISM.

Recoils also decrease the typical total mass of MBH mergers. The total mass of MBH mergers tends to be lower for \textsc{Obelisk-Recoil}. This is because more massive MBHs tend to experience more mergers and recoil kicks. This means that their growth via accretion is suppressed, resulting in smaller masses, but also that they are preferentially ejected from galaxies and therefore missing from subsequent mergers.

\begin{figure}
    \includegraphics[width=\columnwidth]{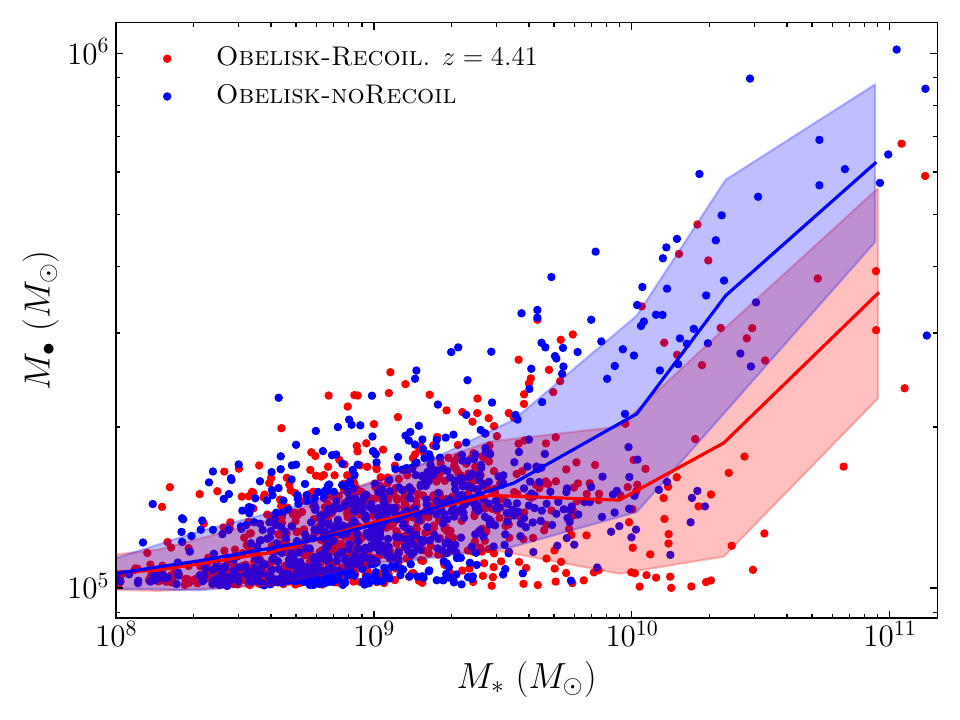}
    \caption{The correlation between the main MBH mass and the host galaxy stellar mass for \textsc{Obelisk-Recoil} (red) and \textsc{Obelisk-noRecoil} (blue). The mean in several galaxy mass bins with the standard deviation is shown in the lines and shaded regions.}
    \label{fig:MBH_Mgal}
\end{figure}

Recoils produce also a noticeable decrease in the normalisation of the $M_\bullet-M_\ast$ relation, shown in Fig.~\ref{fig:MBH_Mgal}. The relative difference between the two simulations remains comparable to the scatter of the relation (and so the error in the mean is smaller than the difference in the mean). We find MBHs close to the seed mass all across the galaxy mass spectrum in both simulations, but differences appear more clearly at $M_\ast>10^{10}\,\rm \Msun$, where in \textsc{Obelisk-noRecoil} more galaxies host MBHs with mass at least twice the seed mass, although the number of objects at the massive end is low because of the small underlying sample of massive galaxies.

Some of our results are qualitatively confirmed by the simulations in which accretion is boosted by setting artificially $v_\mathrm{rel}=0$ in equation~\ref{eq:BHL}. In this case, recoil also reduces the mass of MBHs and the normalisation of the $M_\bullet-M_\ast$ relation, arguing in favour of the robustness of this result. For more details, we refer the reader to Appendix~\ref{app:vrel0}.

In short, we find that recoils can hinder significantly MBH growth, both by reducing the number of mergers and the accretion rate. Our results agree qualitatively with previous studies using different approaches \citep{Micic2006,Sijacki2009,Sijacki2011,Volonteri2011,Blecha2011,Dunn2020}. This makes it even harder to explain the efficient growth of the MBH population inferred from the large population of very massive and active MBHs detected by JWST \citep{Harikane2023,Maiolino2023, Matthee2024,Greene2024}, unless MBHs do not experience mergers in low-mass galaxies.

\subsection{Recoil velocities as a function of MBH evolution}

Physically, the recoil velocity is a function of the spin magnitudes and directions, the direction of the orbital angular momentum and the mass ratio. Typically, higher recoil velocities are sourced by mergers with higher spin magnitudes, and higher degrees of misalignment between the two spin directions and between the spin directions and that of the orbital angular momentum. For non-zero spins, the recoil velocity tends to be maximal for mass ratios close to unity. Therefore, the magnitude of the recoil velocity is linked to the accretion and merger history of the MBHs, over which the spin and masses are built, and to the dynamics of the MBHs, which set the direction of the orbital angular momentum and determine the population of MBH pairs that undergo all dynamical stages from the galaxy merger to coalescence.

\begin{figure}
    \includegraphics[width=\columnwidth]{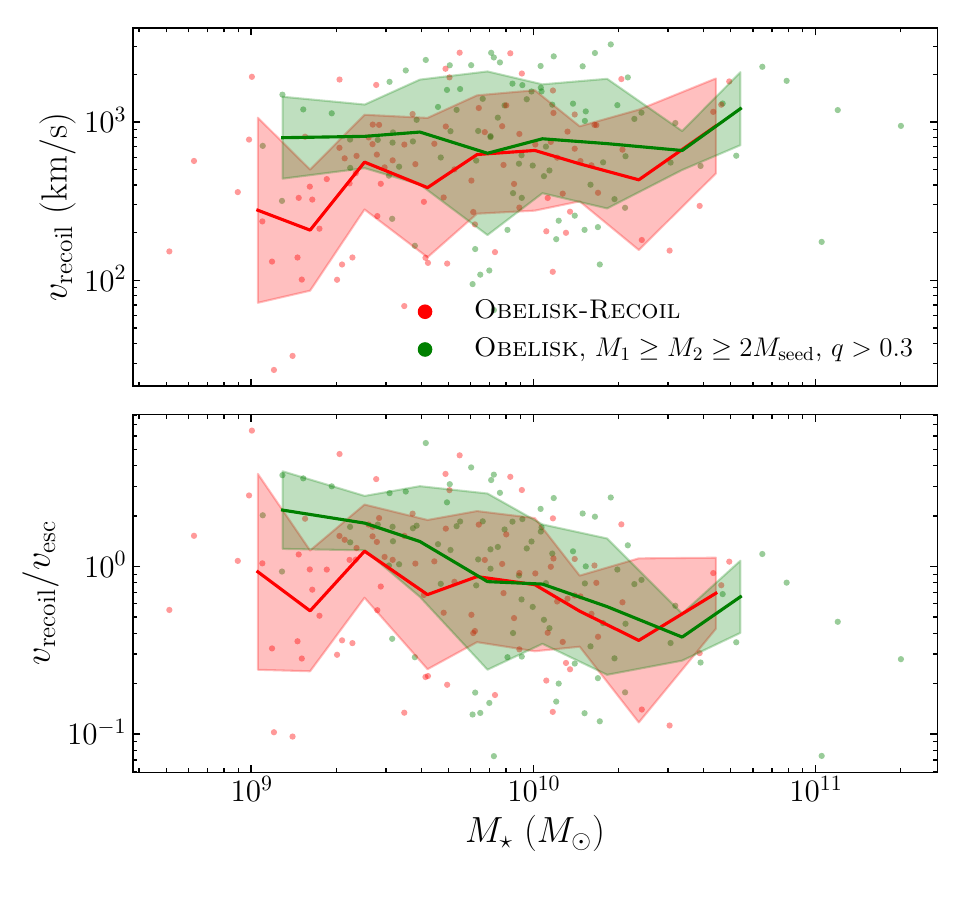}
    \caption{Recoil velocity (top panel) and the ratio of the recoil velocity to the escape velocity of the host halo (bottom panel) as a function of the galaxy stellar mass, for \textsc{Obelisk-Recoil} (red) and \textsc{Obelisk} (green). The recoil kick magnitudes of \textsc{Obelisk} are calculated in post-processing. The mean in several galaxy mass bins with the standard deviation is shown in the lines and shaded regions.}
    \label{fig:recoil_mgal}
\end{figure}

Figure~\ref{fig:recoil_mgal} shows the magnitude of the recoil velocity, $v_\mathrm{recoil}$ and the ratio between the recoil velocity and the escape velocity $v_\mathrm{recoil}/v_\mathrm{esc}$ as a function of the host galaxy mass. This is shown for all the mergers in \textsc{Obelisk-Recoil}. To enlarge the parameter space, we also plot the mergers in the \textsc{Obelisk} simulation, which reaches lower redshift and larger MBH and galaxy masses. For  \textsc{Obelisk} we calculate recoil kick magnitudes in post-processing. For \textsc{Obelisk}, we select major mergers with $q>0.3$ since below this mass ratio the recoil kicks get arbitrarily small and also mergers that fulfill $M_2>2M_\mathrm{seed}$ and so the MBH has erased the model-dependent initial conditions. In this way, we aim to encompass two opposite physical situations -- the MBHs in \textsc{Obelisk-Recoil} have more realistic dynamics and less growth, while the MBHs in \textsc{Obelisk} are dynamically coupled to the gas and experience more efficient growth (and therefore spin coupling to the angular momentum of the gas). That is, \textsc{Obelisk-Recoil} corresponds to more random alignment and \textsc{Obelisk} to more efficient alignment through accretion.

We find that MBHs can receive large recoil kicks after a major merger across a wide range of galaxy masses, MBH masses and accretion rates (MBH masses and accretion rates not shown in the figure). The correlation of $v_\mathrm{recoil}$ with these parameters linked to the MBH evolution is generally weak. We find this to be the case in all our simulations -- \textsc{Obelisk-Recoil}, the simulation with the \textsc{Obelisk-Recoil} model but more efficient accretion (see Appendix~\ref{app:vrel0}), and \textsc{Obelisk}, which suggests that our result is robust for against several accretion models.

The amount of gas during a galaxy merger could have an impact on recoil kicks -- typically, `wet' gas-rich mergers are associated with high spins and aligned configurations as the MBHs couple to the dynamics and angular momentum of the surrounding gas, while `dry' gas-poor mergers are associated with random configurations \citep[][]{Bogdanovic2007,Volonteri2007,Volonteri2010,Blecha2016,Sayeb2021}. 
We find that in our simulation the distinction between high and low accretion rate mergers (relative to Eddington) is not sufficient to drive a bimodality in the magnitude of the kicks. We find only a very weak correlation between the accretion rate before the merger and the alignment of the orbital and spin angular momenta with the surrounding gas.

\begin{figure*}
    \includegraphics[width=\textwidth]{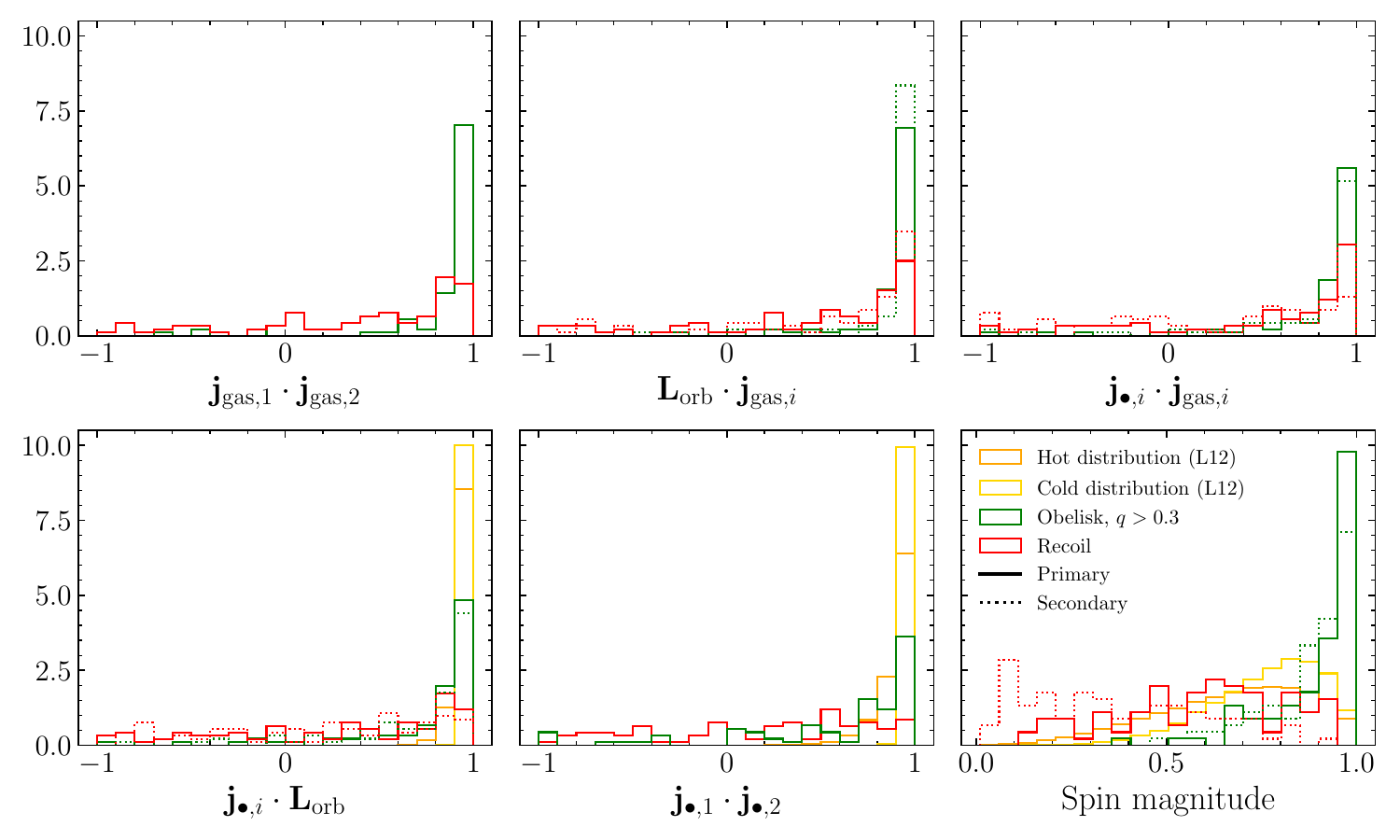}
    \caption{Distribution of the cosine of the angle in the time step before the merger between: the gas angular momenta around each MBH (top left), the orbital angular momentum and the gas angular momenta (top middle), the MBH spins and the gas angular momenta (top right), the MBH spins and the orbital angular momentum (bottom left) and the two MBH spins (bottom middle). The MBH spin magnitudes are shown in the bottom right panel. The distributions are shown for \textsc{Obelisk-Recoil} (red), \textsc{Obelisk} (green), and the `hot' and `cold' disc models in the circumbinary disc simulations of \citet{Dotti2010} (orange and yellow). If both the quantities of the primary and the secondary are shown separately, the primary is shown in solid lines and the secondary in dotted lines.}
    \label{fig:recoil_orientations}
\end{figure*}

Firstly, the dynamical time scale of the MBH merger might not be long enough to align the angular momenta \citep[e.g.][]{Gerosa2015}. However, even if the time scale is long enough, the angular momentum of the nuclear region need not be coherent over time. For low-mass galaxies ($M_\ast\lesssim10^9\,\si{\solarmass}$), strong SN feedback prevents coherent rotationally-supported structures from forming and the MBHs to accrete efficiently \citep{Dubois2014a,Dubois2015,Habouzit2017,Bower2017,Angles-Alcazar2017,Peirani2024,Beckmann2024}, favouring random spins. In higher-mass galaxies ($M_\ast\gtrsim10^9\,\si{\solarmass}$), which are the hosts of most of the mergers in our samples, coherent rotationally-supported gas can produce efficient mass and spin growth. However, the gas angular momentum is rarely completely coherent, there are generally small fluctuations in the direction over short time scales at high accretion rates (e.g. see figures 1 and 2 in \citealp{DongPaez2023a}). These fluctuations can prevent the high pre-merger alignment of angular momenta needed for low recoil kicks. The bottom-centre panel in Fig.~\ref{fig:recoil_orientations} shows the distribution of the alignment between the spins, showing that even if there is a preference for alignment in the simulations, most mergers do not have perfectly aligned spins. The time scale for spin alignment (equation~\ref{eq:t_align}) is typically shorter than that for gas dynamical friction, which means that on top of the misalignment between the spins, the spins will also tend to be misaligned with the orbital angular momentum (centre-left panel, Fig.~\ref{fig:recoil_orientations}). Small misalignments and large spins can still yield large recoil kicks. For example, \citet{Lousto2012} show that rapidly-spinning MBHs misaligned by just $15^\circ$ with the orbital angular momentum and opposite in the azimuthal plane can produce recoil kicks over $1000\,\si{\kilo\meter\per\second}$. In section~\ref{sec:dynamics} we discuss an example of a merger embedded in a rotationally supported disc that nonetheless receives a large recoil kick.

We note that here we are considering numerical mergers, that is, we are ignoring the sub-grid dynamical delay from tens of pc scales to coalescence and the additional alignment that could take place during this evolution. However, if the angular momentum of the circumnuclear discs around bound binaries is preserved from larger scales, the fluctuations in the angular momentum at such scales should also affect the smaller unresolved scales. Furthermore, the timescale for alignment may be longer than the binary evolution timescales \citep{Lodato2013} and idealised high-resolution simulations of MBHs in a circumnuclear disc by \citet{Dotti2010} show that, even if the angular momentum of the circumnuclear disc is coherent, the MBH spins need not fully align with the orbital angular momentum. 

Even if large kicks are produced across all galaxy masses in our sample, the escape velocity increases with the galaxy mass. The ratio $v_\mathrm{recoil}/v_\mathrm{esc}$ of major mergers (Fig.~\ref{fig:recoil_mgal}) decreases slightly with increasing galaxy mass, meaning that in more massive galaxies recoil kicks for major mergers can have a somewhat more moderate but still significant effect.

Figure~\ref{fig:recoil_orientations} investigates further the typical distribution of the spin magnitudes and the relative orientations between the orbital, MBH spin and the angular momentum of the gas around each MBH. The orbital angular momentum and the individual MBH spins tend to be aligned with the ambient gas (top panels). The alignment of each spin with respect to the orbital angular momentum, and the relative alignment between the two MBH spins, which are the terms that enter the calculation of $v_\mathrm{recoil}$, tend to have a somewhat poorer alignment (lower left and middle panels).

Even in the optimistic model of efficient dynamical (and hence angular momentum) coupling to the gas in \textsc{Obelisk} the alignment is high but not perfect. The alignment is lower than in the idealised high-resolution simulations of lower scales (circumbinary discs) by \citet{Dotti2010}. We calculated the distributions by drawing random spins and spin vectors from the analytical fits from \citep{Lousto2012} to the results of \citep{Dotti2010} in the `hot' and `cold'. This is labelled as `L12' in the figure.
Finally, in the bottom-right panels we show that in \textsc{Obelisk}, where MBHs grow efficiently, both primaries and secondaries tend to have spin magnitudes close to maximal. As discussed above, this combination of large spin magnitudes and small misalignment can lead to large recoil velocities with $v_\mathrm{recoil}\gtrsim10^3\,\si{\kilo\meter\per\second}$. The spin magnitudes in \textsc{Obelisk-Recoil} are lower, but alignment is also poorer, as evident comparing the green and red histograms in the lower panels of Fig.~\ref{fig:recoil_orientations}, resulting in similar kick velocities in the two simulations as reported in Fig.~\ref{fig:recoil_mgal}. 

Overall, we find that high recoil velocities can occur in galaxies and MBHs all across the spectrum in masses and accretion rates found in the simulations. We explored two simulations with very different efficiencies in the coupling of MBH angular momenta with the ambient gas. Even if the coupling is efficient and the nuclear gas is rotationally supported there can be small degrees of misalignment that are enough to still produce a high fraction of high recoil velocities. This result is different from simplified models that assume highly aligned configurations \citep[][]{Bogdanovic2007,Volonteri2007,Volonteri2010,Blecha2016,Dunn2020,Sayeb2021}. \citet{Blecha2016} and find that large kicks imply optimistic prospects for the detection of recoiling MBHs as offset AGN. 

\section{Dynamics of recoiling MBHs}
\label{sec:dynamics}

Now we turn to study the orbits of recoiling MBHs. Since the MBHs in \textsc{Obelisk-Recoil} experience limited growth, the sample is composed mostly of major mergers. This makes it suitable for investigating the effect of large recoil kicks with a good statistical sample.

\subsection{Do recoiling MBHs escape their host galaxies?}
\label{subsubsec:escape}

Recoil velocities can be quite large, often exceeding the escape velocity of most MBH merger hosts. Early studies of the cosmological impact of GW recoil using semi-analytic models operated under the assumption that MBHs escape their host galaxies if $v_\mathrm{recoil}>v_\mathrm{esc}$ and remain in the galaxy otherwise \citep[e.g.][]{Haiman2004,Volonteri2005,Schnittman2007}. However, the fate of the MBH does not depend only on the depth of the gravitational well. Dynamical friction, especially from the high stellar densities in galactic cores \citep[e.g.][]{Madau2004}, can remove significant energy from the orbits in the early stages of the recoiling trajectory, and at the subsequent pericentres if the orbit remains sufficiently radial. Moreover, realistic galaxies can be irregular and inhomogeneous -- MBHs might scatter or interact with clumps, satellite galaxies, and other features \citep[e.g.][]{Fiacconi2013,Pfister2019}. The halo also grows and increases its escape velocity with time \citep{Choksi2017}. This problem can only be tackled in its full complexity by using cosmological simulations.

\begin{figure}
    \includegraphics[width=\columnwidth]{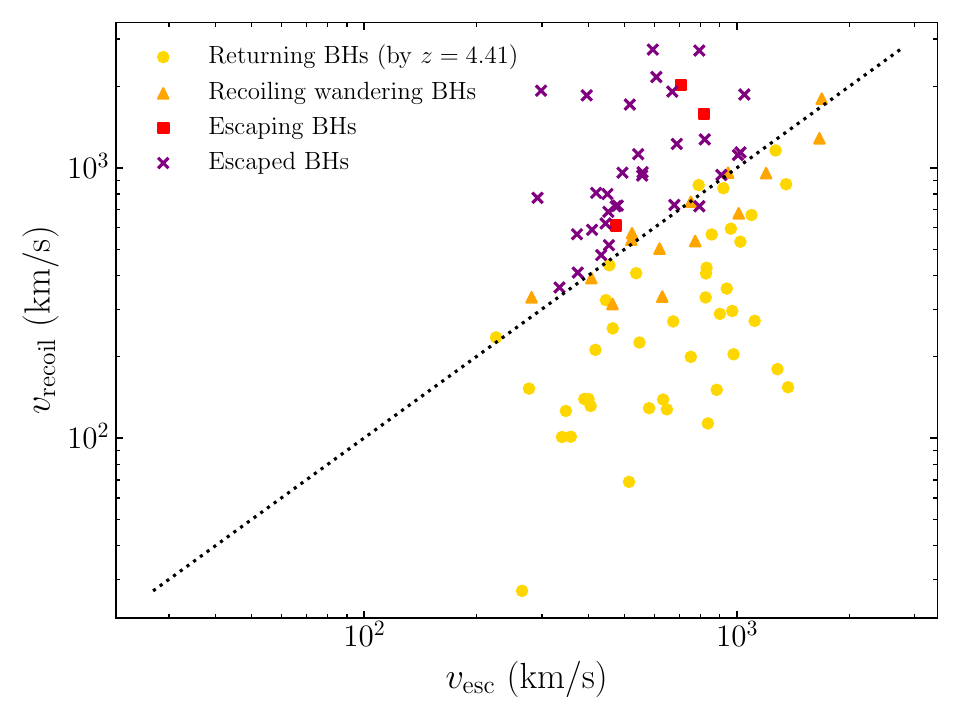}
    \caption{Recoil velocity as a function of the escape velocity of the host for the mergers in \textsc{Obelisk-Recoil}. MBHs that return to $R_{50}$ are shown in yellow points, wandering MBHs are shown in orange triangles, MBHs that are on their way to escaping are shown in red squares, and MBHs that escape their haloes are shown in purple crosses (see table~\ref{tab:dynamical_classification} for a summary of the definitions). The dotted line is the locus where $v_\mathrm{recoil}=v_\mathrm{esc}$. Two outliers caused by spurious measurements have been removed.}
    \label{fig:vrecoil_vesc}
\end{figure}

We find that, in agreement with simplified prescriptions, MBHs tend to escape the halo rapidly if $v_\mathrm{recoil}>v_\mathrm{esc}$ and stay or return rapidly to the galactic centre otherwise. In Fig.~\ref{fig:vrecoil_vesc}, we show the recoil velocity as a function of the escape velocity of the host galaxy. The markers and colours indicate whether the MBH returns to $R_{50}$ before the end of the simulation, stays within the halo as a wandering MBH or escapes the halo. As physically expected, MBHs with $v_\mathrm{recoil}<v_\mathrm{esc}$ do not escape the galaxy. Above $v_\mathrm{esc}$, recoiling MBHs escape their host halo, since dynamical friction is less efficient for fast MBHs that escape the dense regions of the galaxy \citep{Choksi2017}. In total, 37 out of the 92 mergers escape their host haloes.

\begin{figure}
    \includegraphics[width=\columnwidth]{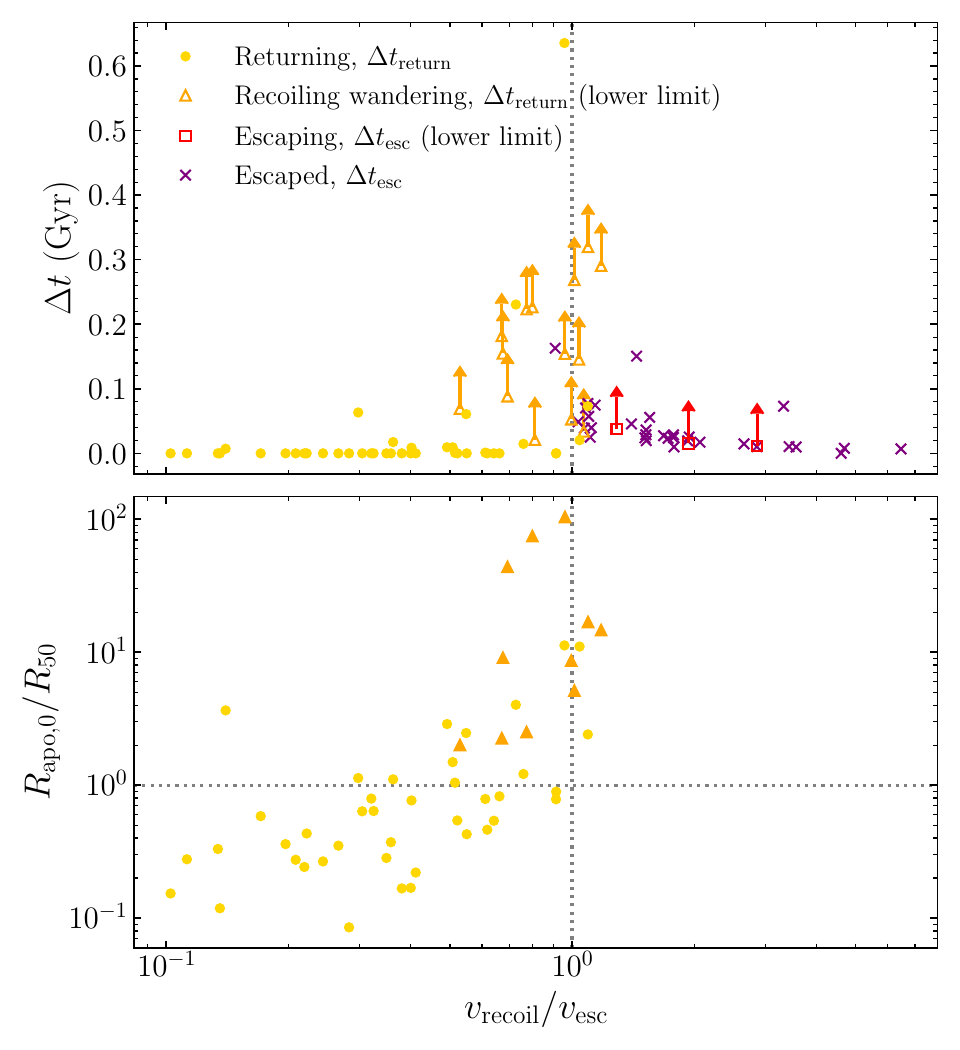}
    \caption{\textbf{Top panel}: the time to return to $R_{50}$, $t_\mathrm{return}$, (for returning, in yellow points, and wandering MBHs, in orange triangles) and or the time to escape the halo, $t_\mathrm{esc}$, (for escaping, in red squares, and escaped MBHs, in purple crosses) as a function of the ratio $v_\mathrm{recoil}/v_\mathrm{esc}$. For wandering and escaping MBHs the time shown is just the time from the recoil kick to the end of the simulation, and should be taken as a lower limit to $t_\mathrm{return}$ or $t_\mathrm{esc}$. \textbf{Bottom panel}: the radius of the first apocentre of the orbit, in units of $R_{50}$ as a function of $v_\mathrm{recoil}/v_\mathrm{esc}$.}
    \label{fig:vrecoil_treturn}
\end{figure}

The time taken for the MBH to return to the centre or to escape behaves asymptotically around $v_{\rm recoil}/v_{\rm esc}\sim1$. Figure~\ref{fig:vrecoil_treturn} shows the time taken to return to $R_{50}$ (a lower limit in the case of wandering MBHs that have not returned by the end of the simulation) or the time to escape the halo (a lower limit in the case of clearly unbound MBHs that have not yet escaped). We observe that for $v_\mathrm{recoil}/v_\mathrm{esc}<0.5$, $t_\mathrm{return}$ is very small or zero -- the MBH never escapes $R_{50}$ or returns in only a few orbits. In the opposite region of the parameter space, for $v_\mathrm{recoil}/v_\mathrm{esc}>2$, the MBHs escape quickly, typically in less than 50 Myr.

Longer return or escape times can only be obtained in a `resonant' regime spanning $0.5\lesssim v_\mathrm{recoil}/v_\mathrm{esc}\lesssim2$. Here, the time needed to return to the galaxy or to escape the halo can be quite long, of the order of several hundreds of Myr. A significant fraction of MBHs in this regime do not return to within $R_{50}$ before the end of the simulation and remain as wandering MBHs. There is still a population of MBHs in this regime that return to $R_{50}$ or escape in a short time.

This trend in $t_\mathrm{return}$ is a reflection of another asymptote in the first apocentric radius of the orbit ($R_{\mathrm{apo},0}$) as a function of $v_\mathrm{recoil}/v_\mathrm{esc}$. Again, only MBHs with $v_\mathrm{recoil}/v_\mathrm{esc}>0.5$ manage to escape the inner regions of the galaxy. In this regime, some of the MBHs manage to dissipate their energy shortly after the kick and stay close to $R_{50}$, while other MBHs escape $R_{50}$ and have difficulties sinking back inside. This is because the dynamical friction time for low-energy orbits is short since the MBH tends to stay in the higher-density regions inside the galaxy. For high-energy orbits, the time scale is long, especially if the MBH spends most of its time in the outskirts or outside of the galaxy in low-density regions. This resembles the analytical results by \citet{Madau2004} assuming a cored single isothermal sphere and the idealised simulations of galaxy mergers by \citet{Blecha2011} -- the radius of the first apocentre diverges as the kick approaches the escape velocity, and so does the dynamical friction time for the MBH to return to the centre.

\begin{figure}
    \includegraphics[width=0.95\columnwidth]{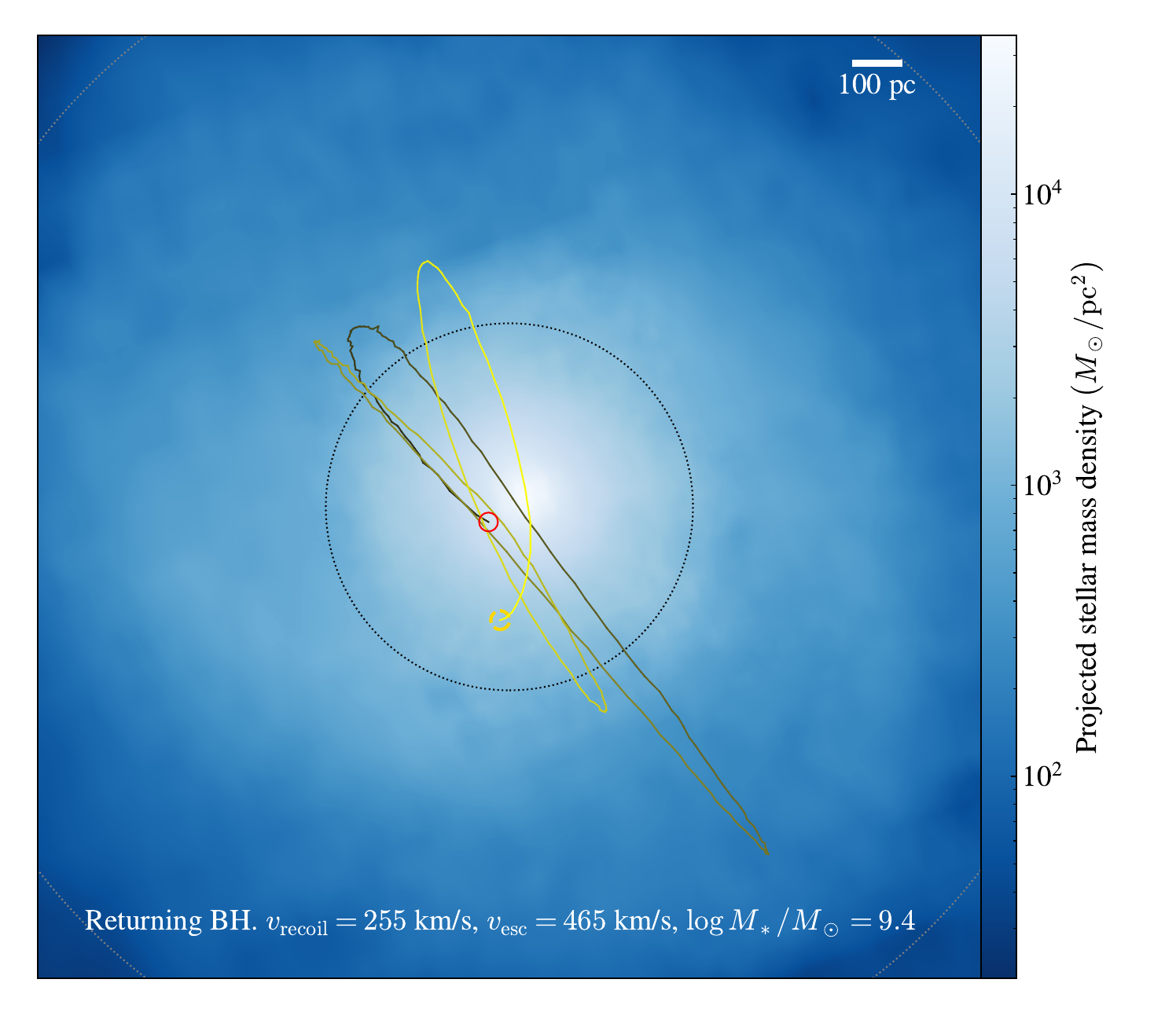}
    \includegraphics[width=0.95\columnwidth]{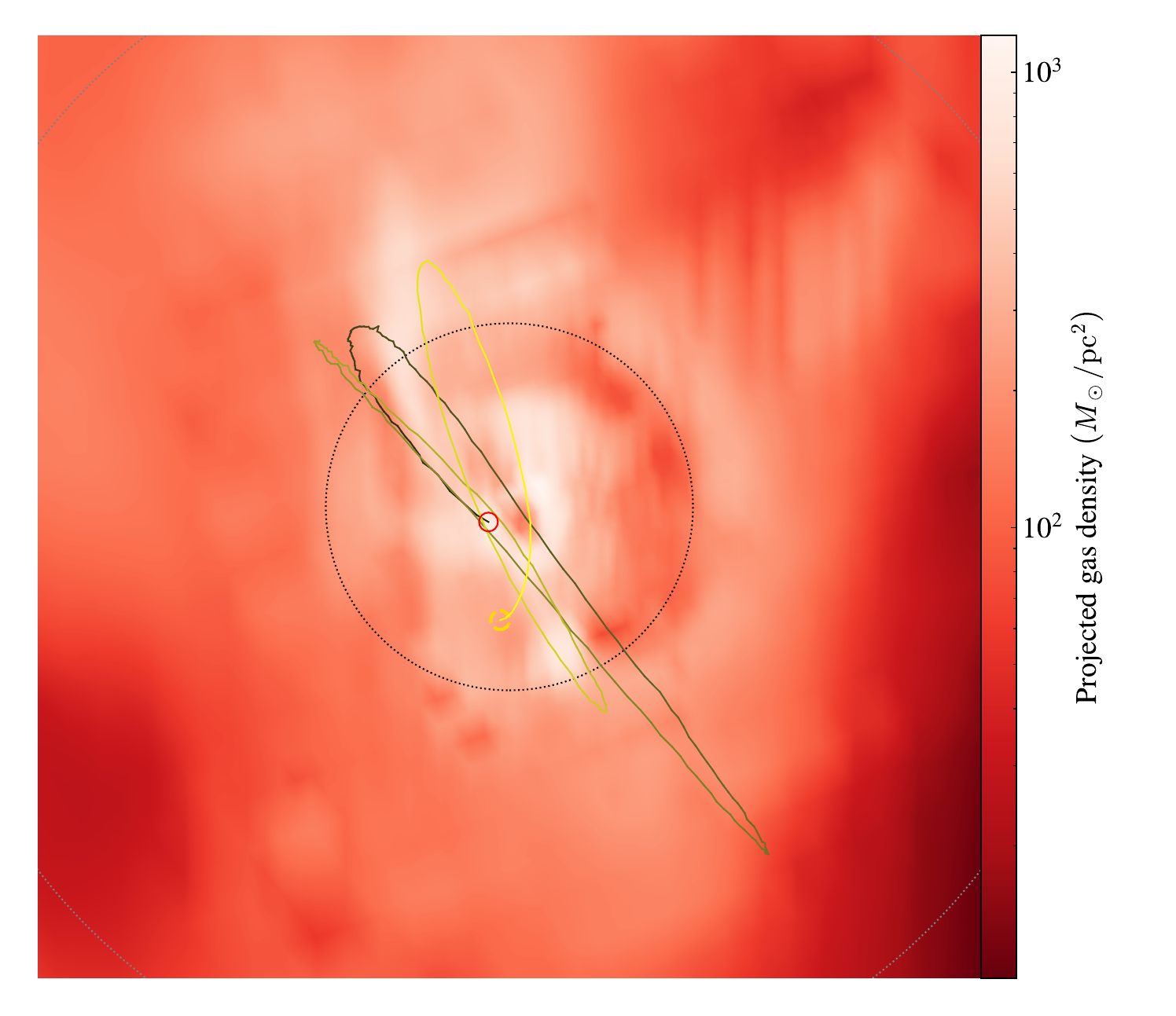}
    \includegraphics[width=0.85\columnwidth]{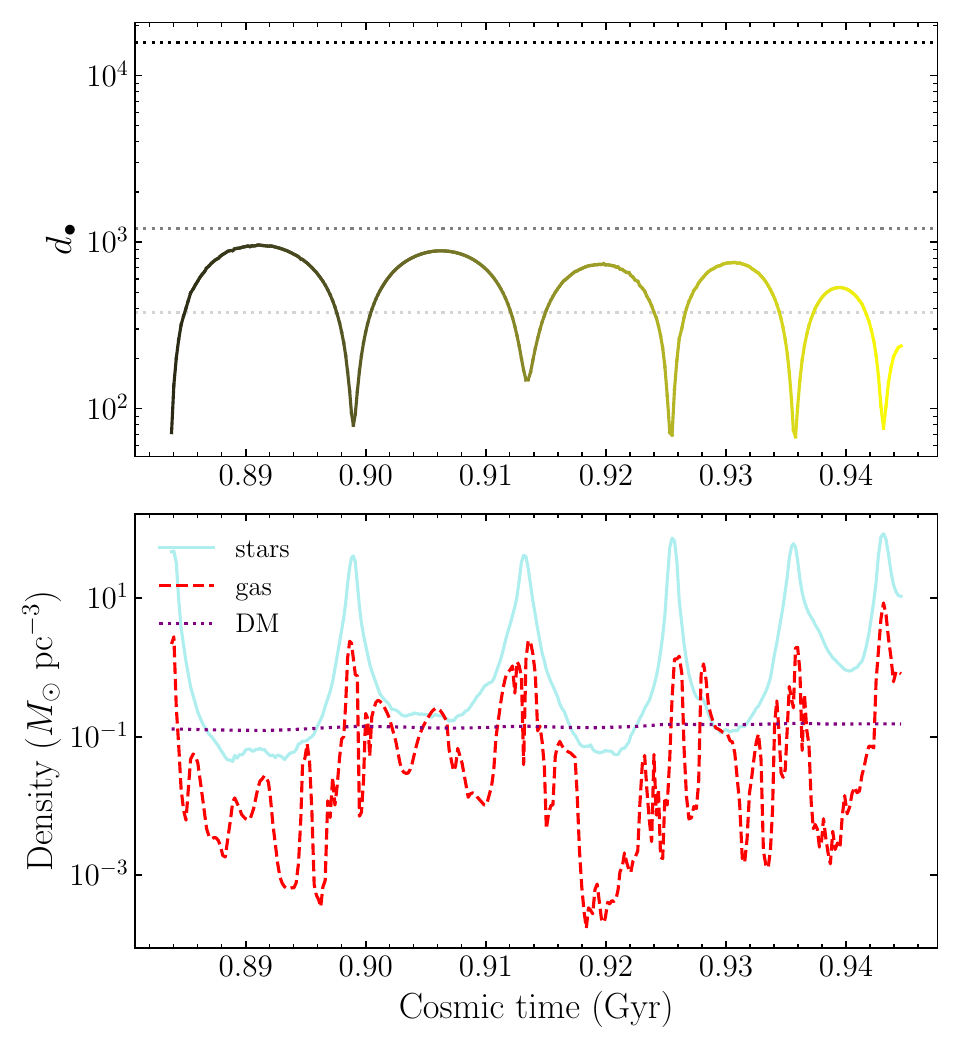}
    \caption{The trajectory of a returning MBH, plotted against the projected stellar and gas density (top and middle panels). The colour lightens with time. The bottom panels show the distance to the centre and the density of stars, gas and DM along the trajectory. Light grey, grey, black dotted horizontal lines denote $R_{50}$, $R_{90}$, $R_\mathrm{vir}$.}
    \label{fig:trajectory_return}
\end{figure}

\begin{figure}
    \includegraphics[width=0.95\columnwidth]{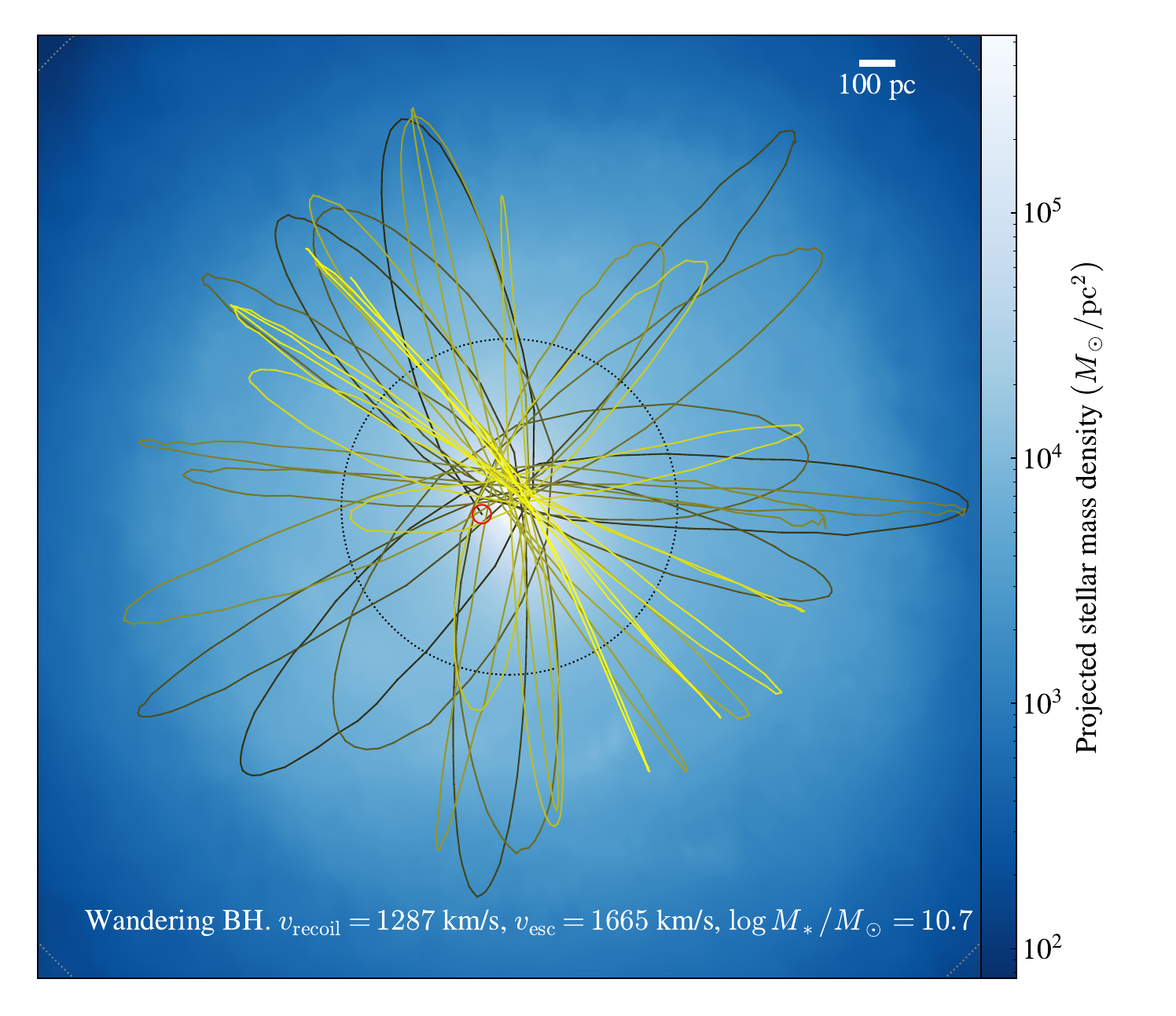}
    \includegraphics[width=0.95\columnwidth]{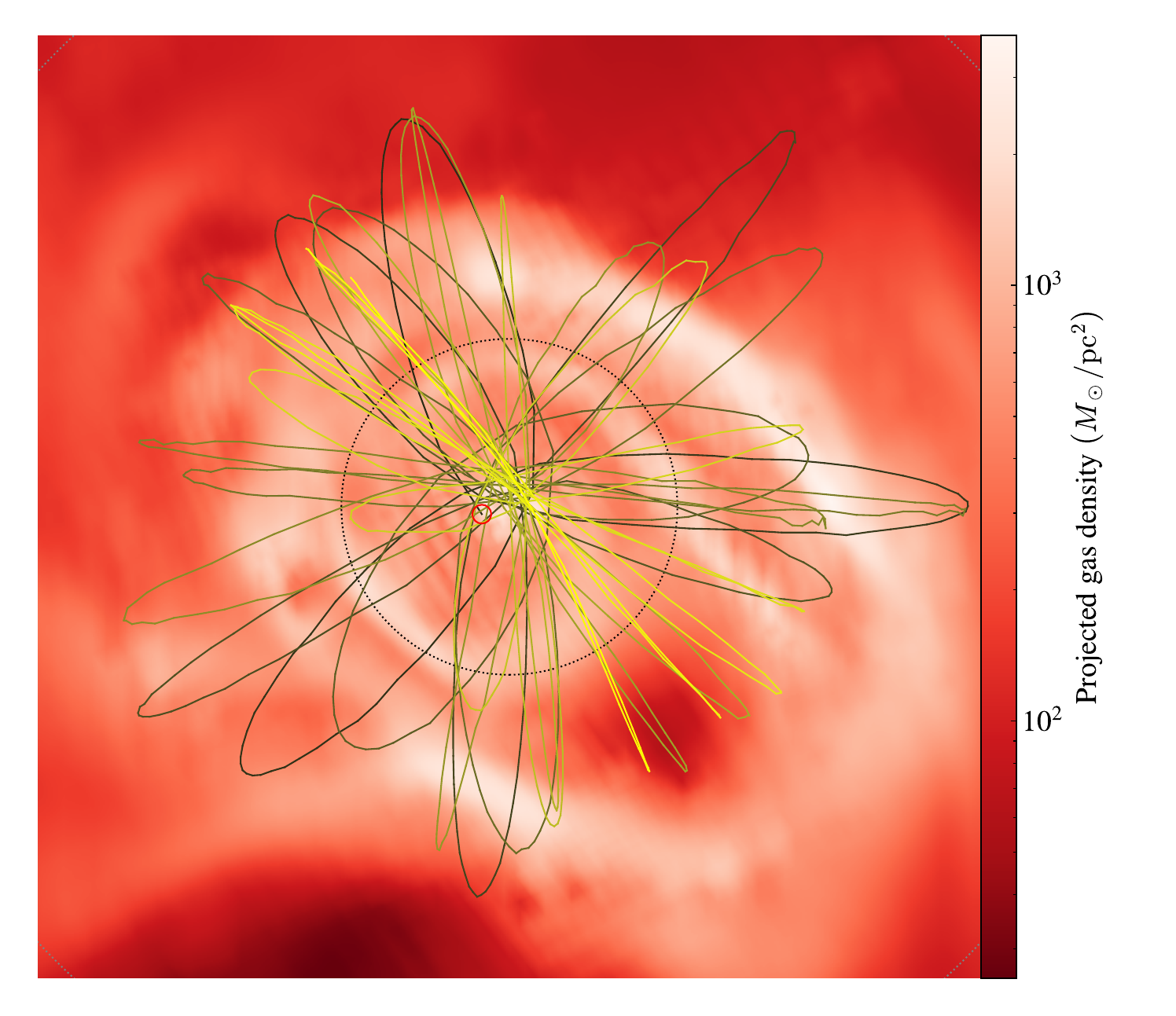}
    \includegraphics[width=0.85\columnwidth]{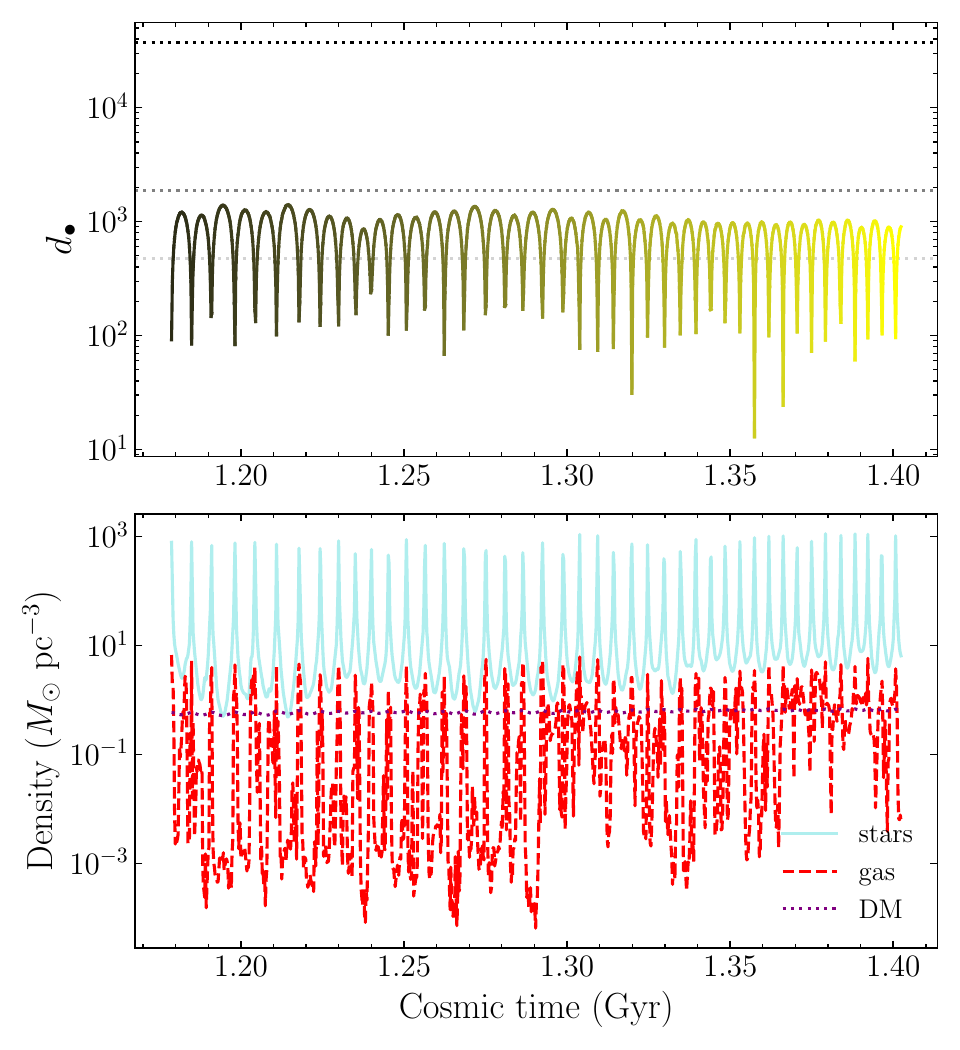}
    \caption{Similar to Fig.~\ref{fig:trajectory_return} but for a recoiling wandering MBH.}
    \label{fig:trajectory_wandering}
\end{figure}

\begin{figure}
    \includegraphics[width=0.95\columnwidth]{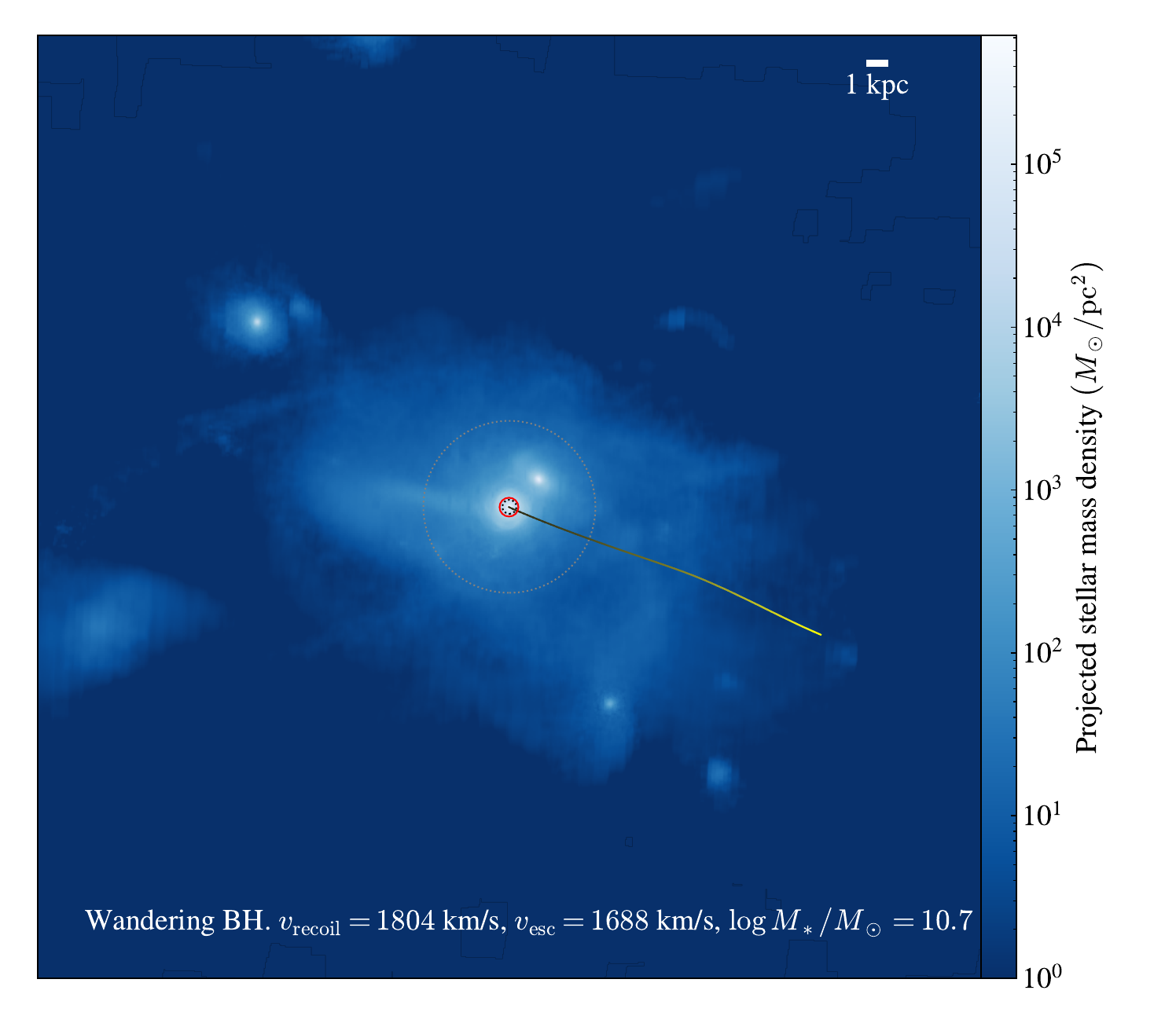}
    \includegraphics[width=0.95\columnwidth]{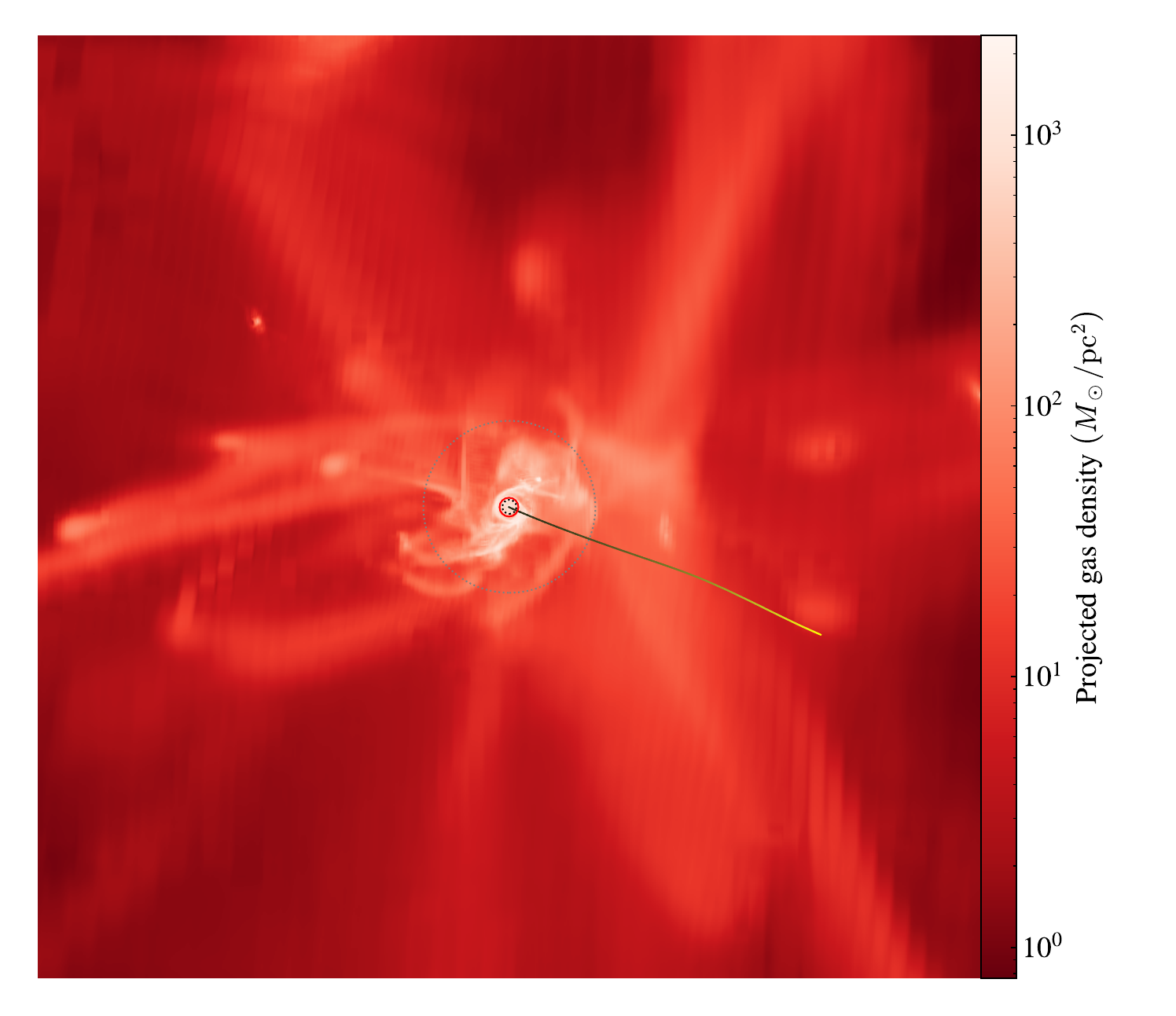}
    \includegraphics[width=0.85\columnwidth]{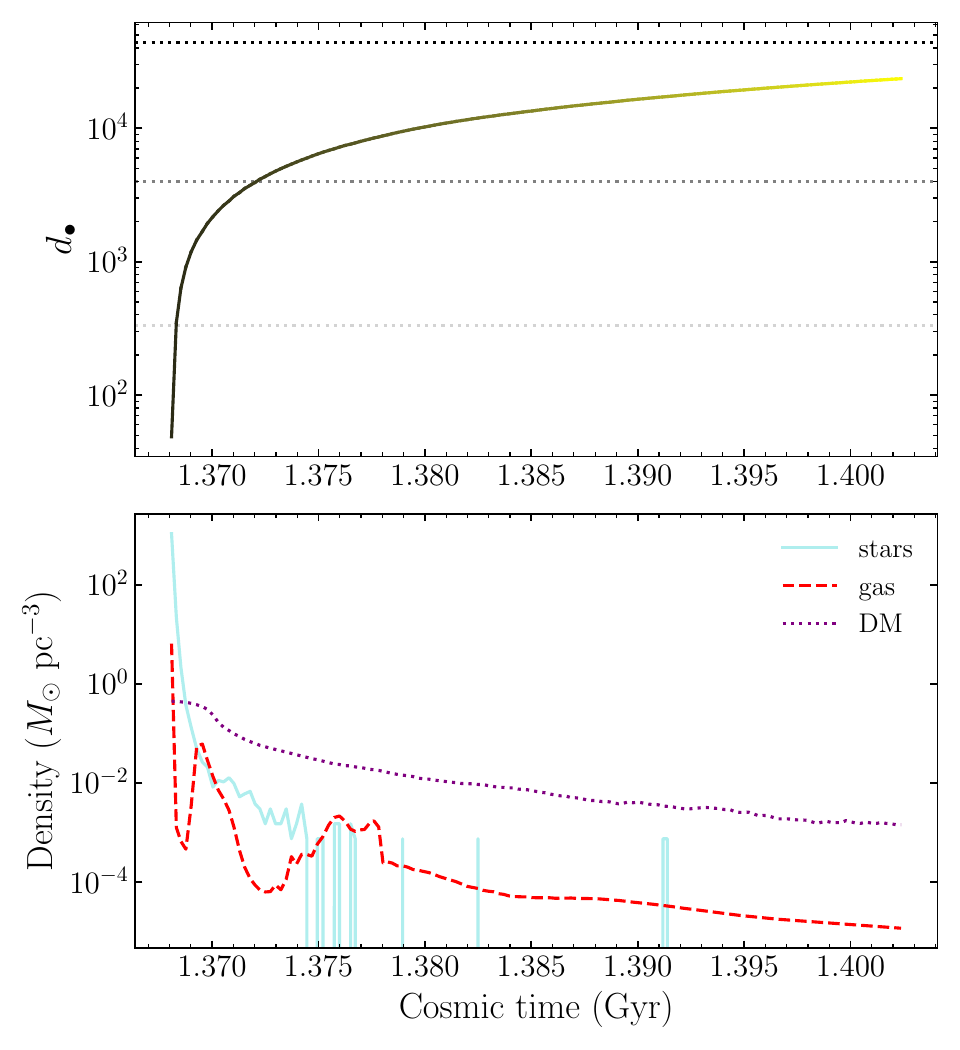}
    \caption{Similar to Fig.~\ref{fig:trajectory_return} but for an MBH as it escapes the host galaxy.}
    \label{fig:trajectory_escape}
\end{figure}

In Figs.~\ref{fig:trajectory_return}, \ref{fig:trajectory_wandering} and \ref{fig:trajectory_escape}, we show examples of trajectories of MBHs with $v_\mathrm{recoil}/v_\mathrm{esc}>0.5$ that return to the centre, wander in the outskirts and escape the halo, respectively. The projections are integrated in the direction of the angular momentum of the galaxy. The galaxies are typically quite compact and dense in the centre, as is common in high-redshift galaxies. The central stellar and gas density tend to be above $10^2$ and $1$ $\si{\solarmass\per\parsec\cubed}$ respectively. The orbit is circularised with time. The MBH in Fig.~\ref{fig:trajectory_return} is ejected outside of the nucleus with $v_{\rm recoil}/v_{\rm esc}=0.55$ but remains inside of the galaxy and decays on a short time scale. As shown in the bottom panel it passes through the central high-density regions, dissipating energy progressively. The MBH in Fig.~\ref{fig:trajectory_wandering} is kicked with $v_{\rm recoil}/v_{\rm esc}=0.77$ to a higher-energy orbit and decays on a longer time scale, remaining as a wandering MBH by the end of the simulation. The orbit is radialised with time. Despite the high central densities of both stars and gas (see bottom panels), the MBH is unable to dissipate energy efficiently. The MBH in Fig.~\ref{fig:trajectory_escape} experiences a larger kick with $v_{\rm recoil}/v_{\rm esc}=1.07$ and escapes its host halo. The densities of gas and stars near the MBH decay rapidly as the MBH exits the galaxy.

\begin{figure}
    \includegraphics[width=\columnwidth]{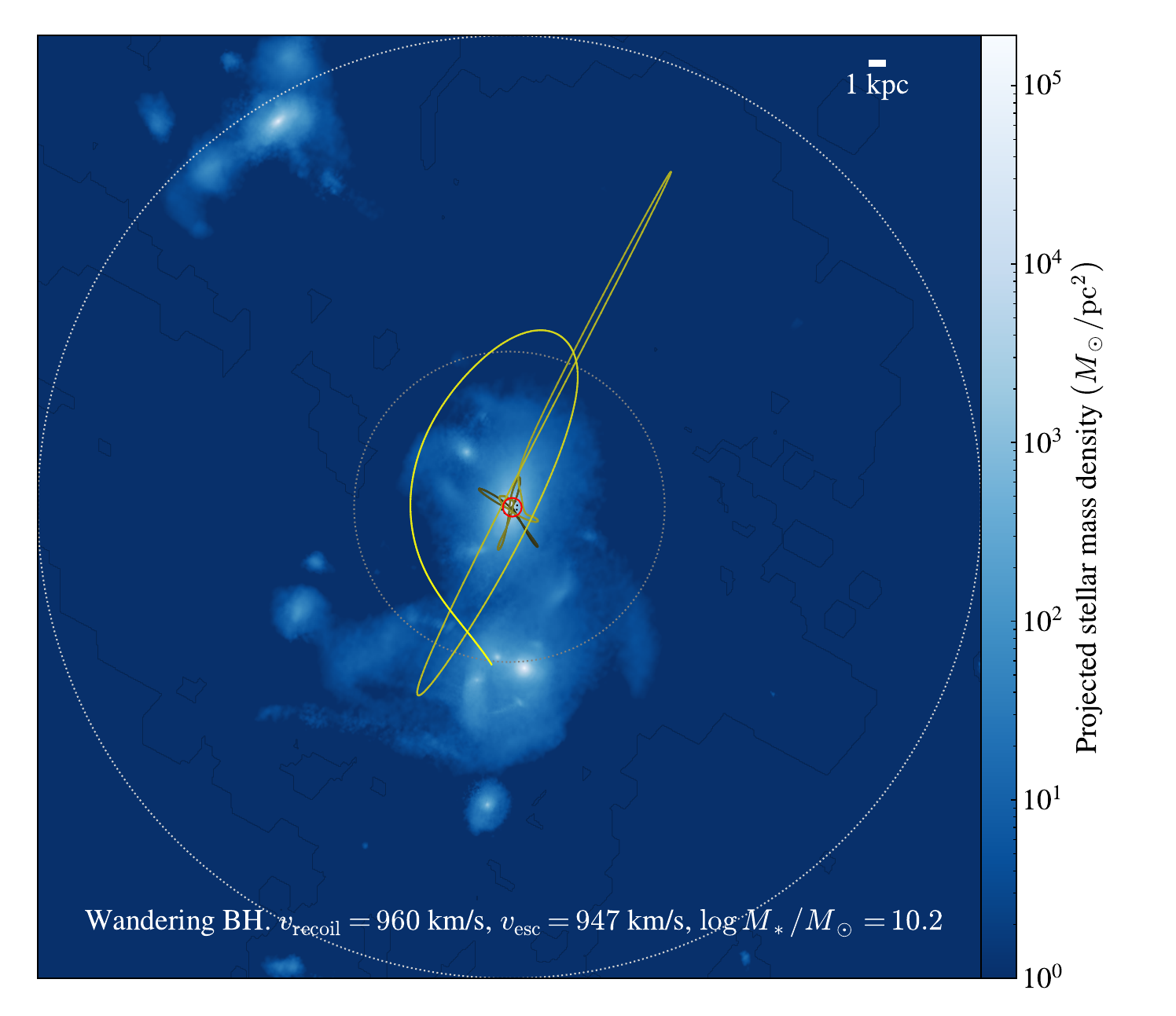}
    \caption{Trajectory of a wandering MBHs that interacts with the halo substructure, plotted against the projected stellar density.}
    \label{fig:trajectory_interaction}
\end{figure}

Dynamical friction is not the only relevant additional effect. The trajectories are often complex. The MBH might interact with clumps and galaxies on its way out -- some recoiling MBHs show signs of gravitational interactions with other structures. An example of this is shown in Fig.~\ref{fig:trajectory_interaction}. The MBH is kicked into a low-energy orbit but scatters later off a sub-halo and gains energy. In other cases, the MBH can even be ejected by such interactions. 
Bound orbits can also radialise or circularise with time, as discussed above. 

In conclusion, in a realistic environment, kicked MBHs tend in most cases to either escape quickly or stay in the centre of the galaxy. There is also a population of MBHs whose fate is difficult to determine without considering the full evolution of the orbit in their complex environment. This is especially the case in the dense and irregular environment characteristic of high-redshift galaxies and haloes.

We highlight that many of our results are also applicable to other possible ejection mechanisms of central MBHs. For example, ejections arising from multiple MBH interactions are likely to be astrophysically relevant \citep[e.g.][]{Volonteri2003,Hoffman2007,Bonetti2019,Partmann2023,Sayeb2024}. Finally, we note that our MBHs release little feedback into their environment, which could reduce the effect of dynamical friction or even reverse it \citep{Sijacki2011,Park2017}.

\subsection{Recoiling MBHs and other wandering MBHs}

We have shown that a significant fraction of recoiling MBHs remain as wandering MBHs. The population of MBHs wandering the galaxy or the halo is not limited to recoiling MBHs. Some other wandering MBHs can originate from former central MBHs that have been displaced \citep{Bellovary2021}, or that have merged into another galaxy and have been stripped of their host galaxies, but dynamical friction has not been efficient enough to sink them to the centre of the potential. Here we investigate whether recoiling MBHs can constitute a significant fraction of the wandering MBH population and whether there are any tell-tale kinematic features distinguishing recoiling MBHs.

\begin{figure}
    \includegraphics[width=\columnwidth]{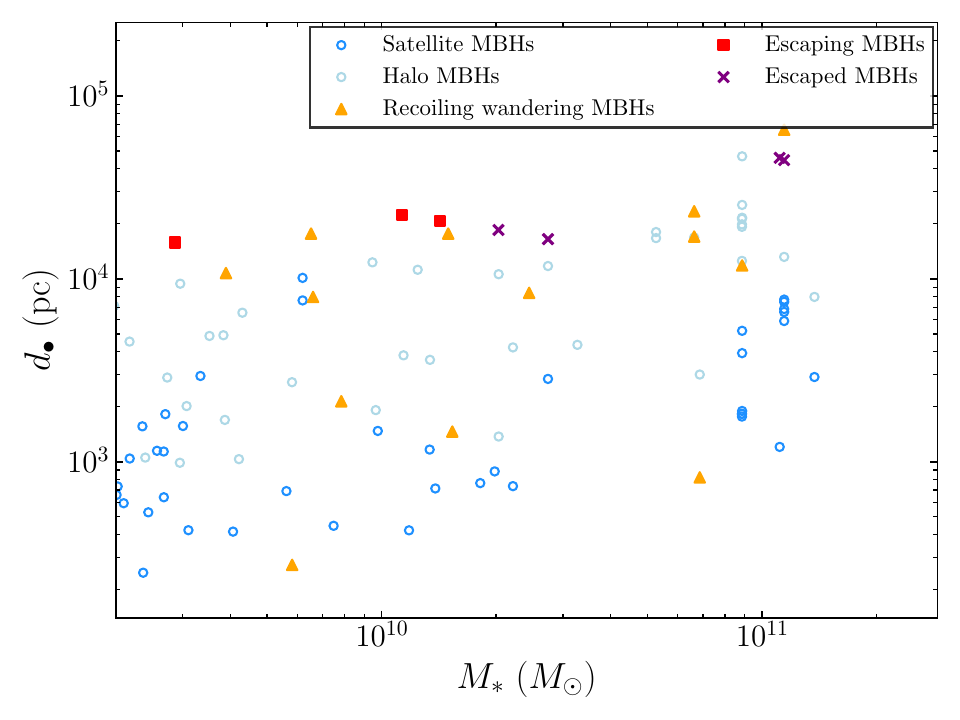}
    \caption{Distance from the galactic centre as a function of the host stellar mass at the last output ($z=4.41$) for MBHs that have previously experienced a recoil kick, which includes wandering (orange triangles), escaping (red squares), and escaped (purple crosses) MBHs. We also show the distribution of MBHs that have not experienced a merger that are main (dark blue empty circles), satellite (blue empty circles), or halo (light blue empty circles) MBHs. See the text for details of the sample selection. The distance corresponding to $4\Delta x$ is denoted with a dashed line.}
    \label{fig:distance_wanderingMBH}
\end{figure}

In our simulations, recoiling MBHs are a significant fraction of the population of wandering MBHs. Figure~\ref{fig:distance_wanderingMBH}, shows the distance of MBHs to the centre of the galaxy as a function of the galaxy mass, for the range in galaxy mass in which mergers occur in our simulation. We observe wandering MBHs at a wide range of distances. In general, we see that recoiling MBHs in halos tend to pile up near $R_\mathrm{vir}$. At lower distances where the dynamical friction times are lower, they represent a smaller fraction compared to wandering MBHs from other origins. In comparison with \citet{Untzaga2024}, we find recoiling MBHs located at smaller distances. We do not consider MBHs outside the virial radius which can be defined in different ways in different models.

As discussed in Fig.~\ref{fig:vrecoil_treturn}, recoiling MBHs are kicked to orbits with a wide range of apocentric distances. Those with small apocentres have a short lifetime as wandering MBHs since they can decay fast through dynamical friction, especially if they remain in radial orbits inside of the galaxy. Those with large apocentres have long dynamical friction times and it is likely to find them as wandering MBHs. Consequently, we find that most recoiling MBHs are either main MBHs that manage to sink back to the centre or MBHs ejected in the resonant regime identified in Fig.~\ref{fig:vrecoil_treturn} in the outskirts of the halo. There are not many recoiling MBHs at intermediated distances, as satellite MBHs inside galaxies. There is also a population of unbound recoiling MBHs on their way to escape the halo. MBHs on high-energy orbits are likely to be found at larger distances, as they spend more time at larger distances where their velocities are smaller. There is also a population (purple crosses) that have exited the halo but then re-enter the same halo or a different one. \textsc{Obelisk-Recoil} models an overdense region, making MBHs more likely to enter another halo upon leaving their original halo.

\begin{figure}
    \includegraphics[width=\columnwidth]{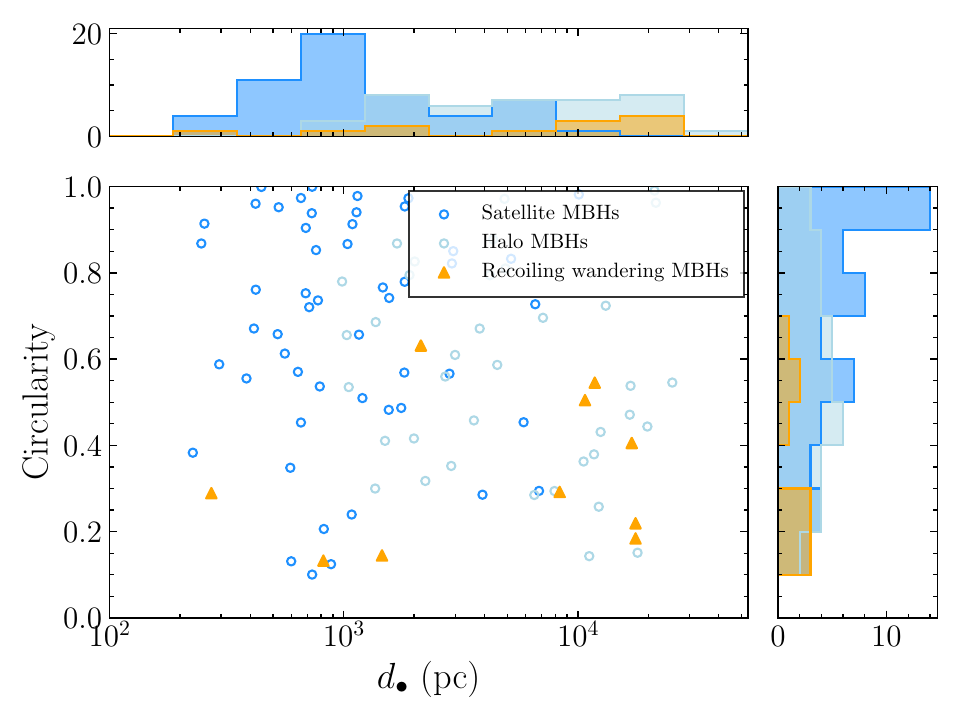}
    \caption{Orbit circularity as a function of the distance from the galactic centre for MBHs associated to galaxies with stellar mass larger than $10^9\,\si{\solarmass}$. The MBH decomposition is similar to that of Fig.~\ref{fig:distance_wanderingMBH}. MBHs that are unbound in a simplified sphericalised potential are not shown. We also exclude recoiling MBHs that have escaped a halo. Only MBHs at distances larger than $2\Delta x$ are shown.}
    \label{fig:circularity}
\end{figure}

Since MBH mergers and recoil kicks occur in the centre of galaxies, one might expect that the orbits of recoiling MBHs are radial. This could be a clear kinematic property to distinguish recoiling MBHs from other kinds of wandering MBHs. However, in our simulations, we find that the picture is more complex. Figure~\ref{fig:circularity} shows the circularity, calculated from the mass distribution in the simulation assuming a spherically symmetric potential (see equation~\ref{eq:circularity}), as a function of the MBH--galaxy distance. The circularity varies from 0 for purely radial orbits to 1 for purely circular orbits. We only show MBHs that are bound in the sphericalised potential. We exclude recoiling MBHs that have escaped a halo.

We observe that many wandering recoiling MBHs retain radial orbits. However, not all of them do. Triaxial or irregular potentials may circularise the orbit \citep{Madau2004,Guedes2009}. Less bound orbits can also be more susceptible to interacting with satellites and being deflected, especially in the violent environments at high redshifts. Even for the recoiling MBHs that preserve radial orbits, this is not a distinctive feature compared to other wandering MBHs. The orbits of the population of other wandering MBHs can also be radial at large distances since satellite galaxies and other objects tend to enter the haloes on radial orbits, as shown also by \citet{Fastidio2024MNRAS}.

\begin{figure}
    \includegraphics[width=\columnwidth]{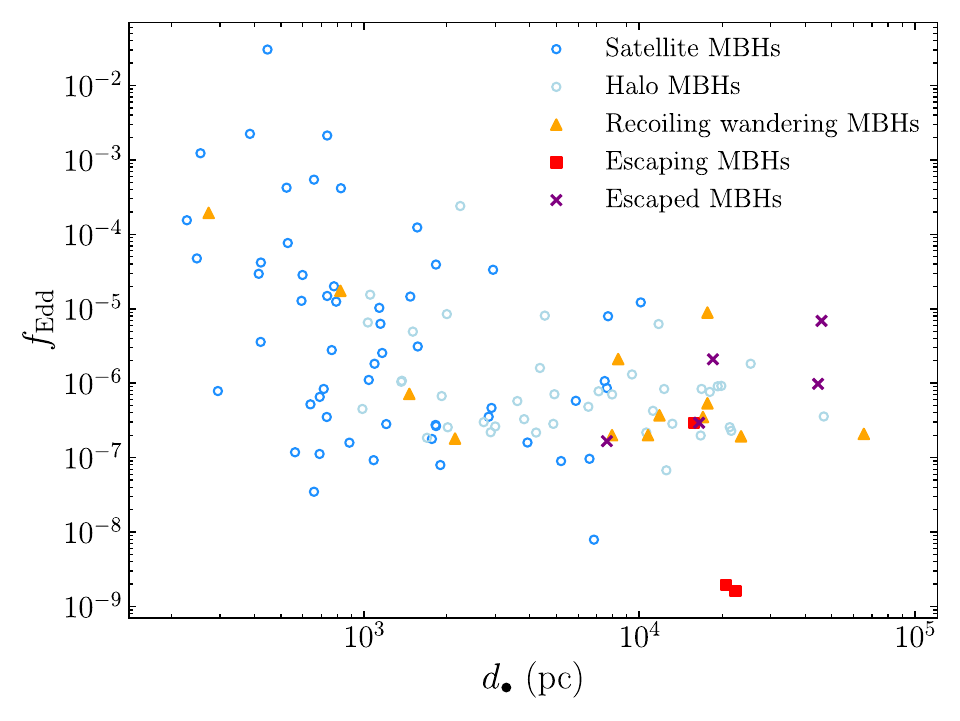}
    \caption{Eddington ratio as a function of the distance from the galactic centre for MBHs that have previously experienced a recoil kick and associated with in galaxies with stellar mass larger than $10^9\,\si{\solarmass}$. The MBH decomposition is similar to that of Fig.~\ref{fig:distance_wanderingMBH}. We exclude all main MBHs and recoiling MBHs that return to the centre of their galaxy.}
    \label{fig:distance_fedd}
\end{figure}

Figure~\ref{fig:distance_fedd} shows the accretion rate as a function of the distance to the centre for the sample of satellite and halo MBHs. The accretion rates drop fast outside of the inner $100\,\si{\parsec}$ with increasing distance. We find again that recoiling MBHs do not show any particular signature in the accretion rate that can distinguish them from other wandering MBHs.

In summary, wandering recoiling MBHs are more likely to be found at large distances, in the outskirts of the halo, compared to other wandering MBHs. Some recoiling MBHs preserve radial orbits, but there are other wandering MBHs with similarly radial orbits at the same distances, making it difficult to discriminate recoiling MBHs just based on dynamics and accretion rate.

\section{Conclusions}
\label{sec:conclusions}

The effect of GW recoil depends on the cosmic evolution and dynamics of MBHs, and vice versa. The dynamics of an MBH after a recoil kick are complex since MBHs inhabit asymmetric and time-varying potentials. Cosmological simulations are required to study this complex problem. We perform a high-resolution simulation (\textsc{Obelisk-Recoil}) of an overdense region down to redshift $z\sim4.4$. The simulation includes a model for GW recoil kicks following MBH mergers coupled to detailed sub-grid models for MBH growth, spin evolution and feedback, dynamical friction, as well as other sub-grid models that are relevant for galaxy formation. We also run a twin simulation ({Obelisk-noRecoil}) without recoil for comparison. We summarise below our main findings.
\begin{itemize}
    \item GW recoil decreases the growth of MBH through accretion (Figs.~\ref{fig:BHgrowth_mergers} and~\ref{fig:fedd_mergers_rec}) and ejects MBHs. This leads to a decrease in the distribution of MBH mass (Fig.~\ref{fig:main_BH_props}), and a decrease in the merger rate. GW recoil also can also decrease the normalisation of the $M_\bullet-M_\ast$ by a factor comparable to the scatter in the relation (Fig.~\ref{fig:MBH_Mgal}).
    \item Nearly-equal mass MBH mergers can produce large kicks ($v_\mathrm{recoil}>1000\,\si{\kilo\meter\per\second}$) which have a large impact on dwarf galaxies, but also on galaxies with stellar masses of $10^{11}\,\si{\solarmass}$ or even higher (Fig.~\ref{fig:recoil_mgal}). Even when MBHs are fed and dynamically coupled to rotationally supported gas, in our simulations we find that the gas is never sufficiently coherent to produce high alignment and low recoil kicks (Fig.~\ref{fig:recoil_orientations}).
    \item After a GW kick, recoiling MBH trajectories behaves asymptotically around $v_\mathrm{recoil}/v_\mathrm{esc}=1$. MBHs with low $v_\mathrm{recoil}/v_\mathrm{esc}$ never escape the centre of the galaxy or return quickly driven by the strong gravity and dynamical friction. MBHs with large $v_\mathrm{recoil}/v_\mathrm{esc}$ escape quickly. Only MBH is the `resonant' regime $0.5\lesssim v_\mathrm{recoil}/v_\mathrm{esc}\lesssim2$ can take longer times ($\sim100\,\si{\mega\year}$) to return or escape (Figs.~\ref{fig:vrecoil_vesc} and~\ref{fig:vrecoil_treturn}).
    \item Some recoiling MBHs remain in the galaxy or the halo as wandering MBHs by the end of the simulation. Recoiling MBHs constitute a significant fraction of the wandering MBH population. Wandering recoiling MBHs tend to be located at large distances outside of the galaxy, where dynamical friction times are longer. Recoiling MBH orbits are complex -- they can change eccentricity with time and get scattered by the substructure in the halo into higher energy orbits or escaping trajectories (Figs.~\ref{fig:trajectory_return},~\ref{fig:trajectory_wandering},~\ref{fig:trajectory_escape}, and~\ref{fig:trajectory_interaction}). Some recoiling MBHs maintain their original radial orbits, but this is not enough to distinguish them dynamically from the global population of wandering MBHs, which can also have radial orbits (Fig.~\ref{fig:circularity}).
\end{itemize}

We conclude that the GW recoil can have a significant effect on the cosmic evolution of MBHs for a wide range of astrophysical environments. It is important to include its effect in cosmological simulations that aim at modeling the growth of MBHs. For example, if one uses the local $M_\bullet-M_\ast$ relation to calibrate an AGN feedback model \citep[as is often done in the field, e.g.][]{Dubois2014} in a simulation that does not account for the suppression of growth from GW recoil, the efficiency of AGN feedback might be overestimated. In the future, we plan to explore the parameter space of MBH evolution models (such as MBH accretion) to understand the effect of MBH evolution on GW recoil, couple GW kicks with the recently developed RAMCOAL model to track subgrid MBH binary dynamics~\citep{Li2024}, and include ejections from triple interactions.

\begin{acknowledgements}
We thank Francisco Rodr\'iguez Montero, Luc Blanchet, Corentin Cadiou, Yosef Zlochower, Carlos O. Lousto, Eugene Vasiliev, Kunyang Li, Alberto Mangiagli, Nimatou Diallo, and David Trestini for useful discussions. This work was made possible by funding from the French National Research Agency (grant ANR-21-CE31-0026, project MBH\_waves). This work was supported by the CNES for the space mission LISA.  This work has made use of the Infinity Cluster hosted by Institut d’Astrophysique de Paris; we thank Stéphane Rouberol for running smoothly this cluster for us. For the purpose of open access, the author has applied a Creative Commons Attribution (CC BY) licence to any Author Accepted Manuscript version arising from this submission.
\end{acknowledgements}


\bibliography{references}

\begin{appendix}

\section{Simulations with $v_\mathrm{rel}=0$}
\label{app:vrel0}

We ran twin simulations of \textsc{Obelisk-Recoil} and \textsc{Obelisk-noRecoil} varying the accretion model, namely setting $v_\mathrm{rel}=0$ in equation~\ref{eq:BHL}. 

\begin{figure}
    \includegraphics[width=\columnwidth]{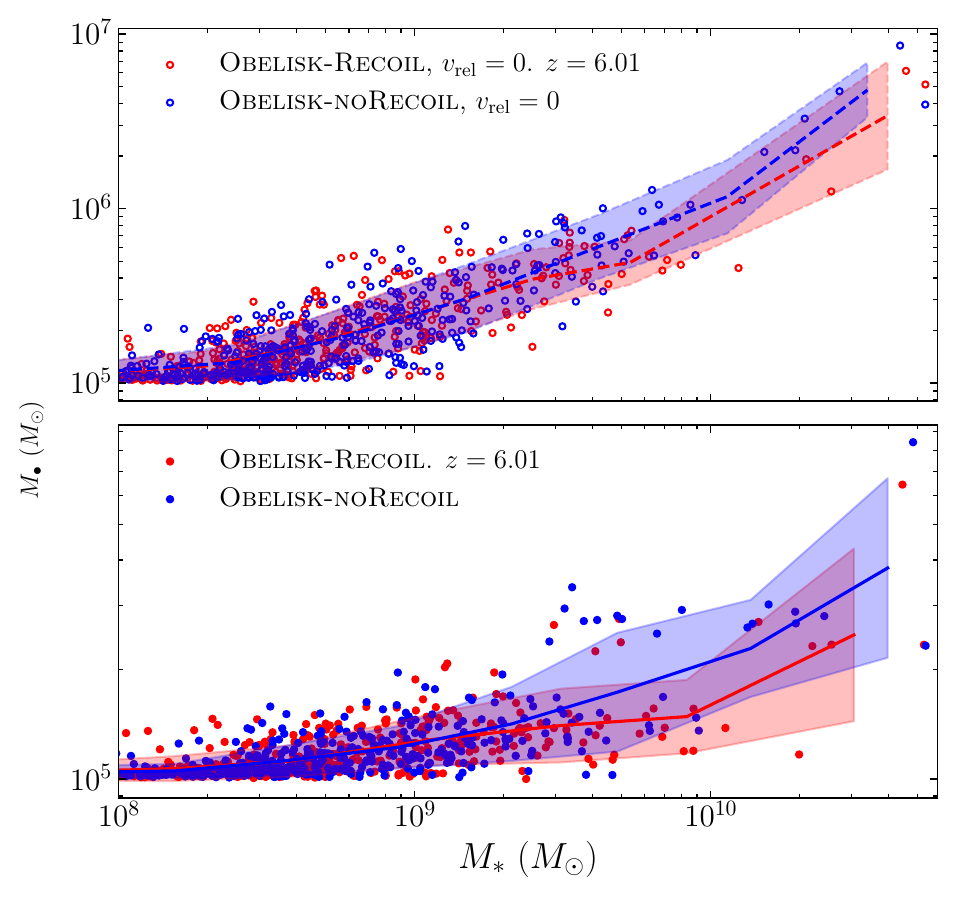}
    \caption{Similar to Fig.~\ref{fig:MBH_Mgal}. The correlation between the main MBH mass and the host galaxy stellar mass for the $v_\mathrm{rel}=0$ versions of \textsc{Obelisk-Recoil} (red) and \textsc{Obelisk-noRecoil} (blue) (top panel) and the fiducial versions (bottom panel) at $z=6.01$.}
    \label{fig:MBH_Mgal_vrel0}
\end{figure}

Figure~\ref{fig:MBH_Mgal_vrel0} shows the $M_\mathrm{\bullet}-M_\ast$ relation for the $v_\mathrm{rel}=0$ and the fiducial sets of simulations. In both cases, we find that the number of the most massive MBHs is reduced in the recoil compared to the no recoil simulations, in both the $v_\mathrm{rel}=0$ and fiducial cases. Similarly, with $v_\mathrm{rel}=0$, the normalisation of the $M_\mathrm{\bullet}-M_\ast$ relation is reduced in the recoil simulations compared to the no recoil simulations.

\begin{figure}
    \includegraphics[width=\columnwidth]{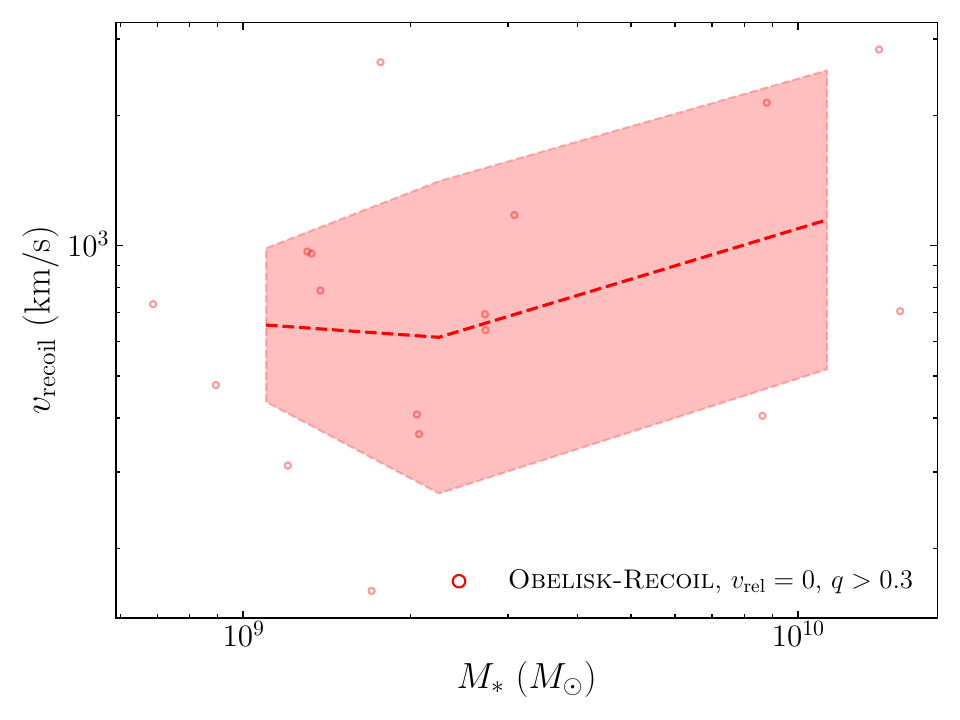}
    \caption{Similar to Fig.~\ref{fig:recoil_mgal}. Recoil velocity as a function of the galaxy stellar mass, for the major mergers in the $v_\mathrm{rel}=0$ version of \textsc{Obelisk-Recoil}.}
    \label{fig:recoil_mgal_vrel0}
\end{figure}

Figure~\ref{fig:recoil_mgal_vrel0} shows the recoil velocity as a function of the host galaxy stellar mass. This confirms the trend found for \textsc{Obelisk-Recoil} and \textsc{Obelisk} -- large recoil kicks can occur in a variety of environments and galaxy masses.

Overall, we find qualitatively similar results when the model is varied to make accretion more efficient. This suggests that the results presented in the main body of the paper are robust.

\end{appendix}

\end{document}